\documentclass[journal]{IEEEtran}
\usepackage{cite}

\ifCLASSINFOpdf
\else
\fi
\usepackage{graphicx}
\usepackage[outdir=./figures/]{epstopdf}
\graphicspath{{./figures/}}
\usepackage{tikz}
\usetikzlibrary{matrix,automata,arrows,arrows.meta}
\usetikzlibrary{shapes,shadows,positioning,calc}
\usepackage{xcolor}

\usepackage{amsmath}
\usepackage{amssymb}
\usepackage{mathtools}
\usepackage{stmaryrd}
\usepackage[binary-units=true]{siunitx}

\usepackage{algorithm}
\usepackage{algpseudocode}

\usepackage{enumerate}
\usepackage{tabularx}
\usepackage{multirow}
\usepackage{makecell}
\usepackage{arydshln}
\usepackage{booktabs}

\usepackage{caption}
\usepackage{subcaption}

\newcommand{\until}{\mathbf{U}}

\newcommand{\always}{\Box}
\newcommand{\eventually}{\lozenge} 

\newcommand{\ltlx}{\text{LTL}}

\newcommand{\win}{\text{Win}}

\newcommand{\pre}{\text{Pre}}

\newcommand{\trace}{\text{Trace}}

\newcommand{\abs}[1]{\left\vert#1\right\vert}
\newcommand{\norm}[1]{\left\Vert#1\right\Vert}
\newcommand{\set}[1]{\left\{#1\right\}}
\newcommand{\Ball}[1]{\mathcal{B}_{#1}}

\newcommand{\outedge}{\text{Out}}

\newcommand{\card}[1]{\abs{#1}}
\newcommand{\interior}[1]{\text{int}({#1})}

\newcommand{\norminf}[1]{\left\| #1 \right\|}

\newcommand{\sv}{\,\vert\;}

\newcommand{\sys}{\mathcal{S}}
\newcommand{\word}{\mathbf{w}}

\newcommand{\Real}{\mathbb R}

\newcommand{\Z}{\mathbb Z}
\newcommand{\Zp}{\mathbb{Z}^+}
\newcommand{\N}{\mathbb{N}}

\newcommand{\M}{\mathbf{M}}
\newcommand{\T}{\mathbf{T}}

\newcommand{\A}{\mathcal{A}}
\renewcommand{\P}{\mathcal{P}}

\newcommand{\ctlr}{\mathcal{C}}

\newcommand{\D}{\mathcal{D}}
\newcommand{\graph}{\mathcal{G}}
\newcommand{\W}{\mathbf{W}}

\newcommand{\x}{\mathbf{x}}

\newcommand{\uu}{\mathbf u}
\newcommand{\dd}{\mathbf d}
\newcommand{\K}{\mathcal{K}}
\newcommand{\X}{\mathcal{X}}
\newcommand{\act}{\mathcal{U}}

\renewcommand{\subset}{\subseteq}

\DeclarePairedDelimiter{\ceil}{\lceil}{\rceil}

\newtheorem{defn}{Definition}
\newtheorem{prop}{Proposition}

\newtheorem{assum}{Assumption}
\newtheorem{rem}{Remark}
\newtheorem{thm}{Theorem}
\newtheorem{lem}{Lemma}
\newtheorem{cor}{Corollary}
\newtheorem{exmp}{Example}

\begin{document}
%
\title{A Specification-Guided Framework for Temporal Logic Control of Nonlinear Systems}
%
%
%

\author{Yinan~Li,
        Zhibing~Sun,
        Jun~Liu,~\IEEEmembership{Senior Member,~IEEE}
        \thanks{This work is supported in part by the NSERC DG, CRC, and ERA programs. Z. Sun and J. Liu are with the Department of Applied Mathematics, University of Waterloo, Waterloo, Ontario, Canada. Y. Li is currently with Clearpath Robotics, Inc. This work was done while she was with the Department of Applied Mathematics, University of Waterloo. {\tt\small yinan.li, zhibing.sun, j.liu@uwaterloo.ca}}%
}

\maketitle

\begin{abstract}
This paper proposes a specification-guided framework for control of nonlinear systems with linear temporal logic (LTL) specifications. In contrast with well-known abstraction-based methods, the proposed framework directly characterizes the winning set, i.e., the set of initial conditions from which a given LTL formula can be realized, over the continuous state space of the system via a monotonic operator. Following this characterization, an algorithm is proposed to practically approximate the operator via an adaptive interval subdivision scheme, which yields a finite-memory control strategy. We show that the proposed algorithm is sound for full LTL specifications, and \emph{robustly complete} for specifications recognizable by deterministic B\"uchi automata (DBA), the latter in the sense that control strategies can be found whenever the given specification can be satisfied with additional bounded disturbances. Without having to compute and store the abstraction and the resulting product system with the DBA, the proposed method is more memory efficient, which is demonstrated by complexity analysis and performance tests. A pre-processing stage is also devised to reduce computational cost via a decomposition of the specification. We show that the proposed method can effectively solve real-world control problems such as jet engine compressor control and motion planning for manipulators and mobile robots.
\end{abstract}


%
\IEEEpeerreviewmaketitle

\section{Introduction}
%
%
%
%

\subsection{Background and motivation}

\IEEEPARstart{T}{he} use of formal methods in control synthesis automates the procedure of control system design and yields provably correct control strategies \cite{BeltaBook2017}. Another benefit is that complex specifications involving discrete logic can be handled, while conventional methods are usually designed for simple control objectives such as stabilization. Such specifications can be described by linear temporal logic (LTL) and are commonly used in motion planning for enforcing reactive missions, task coverage and sequencing \cite{Bhatia2011,Plaku2015}.

Current solutions to a general LTL control synthesis problem are mostly based on discrete abstractions of the original continuous-state system. Algorithms for solving two-player games \cite{Zielonka1998} are then applied to the product system of the abstraction and the deterministic automaton translated from the LTL specification \cite{BeltaBook2017}. For the most part of this paper, we restrict our scope to the LTL formulas that can be translated into deterministic B\"uchi automata (DBA). Many of the specifications in motion planning problems can be expressed by DBA and the fragment of LTL called co-safe LTL (scLTL), which is often used for system safety verification \cite{Kupferman2001}, is also included. Control synthesis is said to be sound and complete if it accurately determines the winning set, i.e., the set of states from which the control specification can be satisfied. Soundness and completeness can be achieved at the abstraction level, but the gap between the continuous-state systems and their finite abstractions leaves the question open: \emph{can LTL/DBA control synthesis be sound and complete under general nonlinear dynamics?}

To bridge the gap, efforts have been made on proposing various system relations \cite{Tabuada2009verification}. Without stability assumptions, we can construct sound finite abstractions by over-approximating (and hence is conservative) the dynamics \cite{ZamaniPMT12,liu2016finite,Reissig2016}, but they are still nonequivalent to the original dynamical systems. A sound and approximately complete abstraction \cite{liu2017robust} may contain an exceptionally large number of states and transitions, because finite abstractions are often obtained by uniformly discretizing the system state space according to the global Lipschitz constant. While controller synthesis can be conducted offline, the prohibitive size of the abstraction will render the synthesized controller impractical due to the significant memory cost for implementing it. This raises another practical question: \textit{can we exploit the structure of the specification to provide a more memory efficient framework for control synthesis?}

Motivated by these questions, this paper proposes a specification-guided framework for temporal logic control of nonlinear systems. In contrast with commonly used abstraction-based methods, where abstraction and synthesis are independent of each other, the current paper characterizes the winning set w.r.t. a given specification directly over the continuous state space of the original system. The B\"uchi automata (BA) translation of the LTL specification is used to guide an adaptive partition scheme during control synthesis. The key to integrating discrete automaton states and the continuous-state space is a monotonic fixed-point operator, which can be practically approximated via the tool of interval analysis \cite{MooreBook66} and an adaptive subdivision scheme. Hence, we are able to provide a practical control synthesis algorithm that terminates in finite time and guarantees a notion of completeness. While fixed-point algorithms for directly solving B\"uchi games are well known \cite{chatterjee2008algorithms,chatterjee2012n} and the earlier work \cite{de2001symbolic} investigated semi-algorithms for solving $\omega$-regular games on infinite-state spaces, to the best knowledge of the authors, there are currently no efficient implementations of such algorithms for direct synthesis of controllers for continuous-state nonlinear systems. Earlier works \cite{Wolff2013icirs,AydinGol2014} also use specification-guided ideas to solve LTL control problems. Our work is different from theirs in that \cite{Wolff2013icirs} only considers sound solutions and relies on solving underlying set-to-set constrained reachability problems by other means, and the method in \cite{AydinGol2014} is restricted to linear or piecewise affine systems. None of \cite{Wolff2013icirs,AydinGol2014} discussed theoretical guarantees of finding controllers. In this paper, we not only provide an effective implementation of a specification-guided control synthesis framework for general nonlinear systems that directly connects the low-level dynamics with high-level specifications, but also prove theoretical guarantees of completeness as detailed below.

\subsection{Main contributions}

\subsubsection{Theoretical contributions} While sound and complete LTL control synthesis algorithms exist for (non-deterministic) finite transition systems (see \cite{Kloetzerhscc08,BeltaBook2017}), whether or not LTL (including DBA) control synthesis can be complete for continuous-state dynamical systems is still an open question. As a main technical contribution, we prove in Theorem \ref{thm:maindba} that DBA control synthesis for continuous-state systems is robustly complete in the sense that control strategies can be found whenever the given specification can be satisfied with additional bounded disturbances.
Furthermore, we show that the proposed DBA control synthesis algorithm can also be used to generate sound solutions to full LTL control synthesis problems.
Invariance is one of the simplest LTL formulas that can be translated to DBA, and the corresponding robustly complete control synthesis algorithm has been presented in \cite{Li2016}. In \cite{Li2020Reachstay}, a robustly complete algorithm is proposed for the reach-and-stay specification ($\eventually\always B$), which is a simple LTL formula that cannot be translated into DBA. The current work provides a general framework for DBA control synthesis problems with theoretical guarantees.

\subsubsection{Practical contributions} We show that the specification-guided framework based on an adaptive partition of the state space is more memory-efficient than abstraction-based methods. In some benchmarking instances, this implementation leads to a 300-fold reduction in memory consumption for synthesis. This is because the proposed method does not pre-compute and store all the transitions of both the abstraction and the product system of the abstraction and the DBA. Having low memory consumption makes it possible to perform DBA control synthesis on a personal computer with limited random access memory (RAM), while the abstraction-based approach can still fail on nodes of high performance computing (HPC) clusters with significant memory capacity. To compare and analyze time and memory performances of the proposed method and abstraction-based methods, we present case studies on Moore-Greitzer engine control, motion planning of manipulators and mobile robots with different specifications, in which different partition precisions are required. Experiments show that the proposed method has an overall better performance than abstraction-based methods when high partition precision is required. It provides a useful alternative to abstraction-based synthesis when low memory consumption is critical for synthesis and implementation of controllers. 

\subsubsection{Heuristic contributions}

Computational complexity is a major concern in general LTL control synthesis. While the restriction to DBA specifications can significantly reduce the complexity, a promising finite abstraction of nonlinear dynamics is usually huge in size, and control synthesis on a product system would still be intractable because the number of states is the multiplication of the sizes of both the abstraction and the DBA. In \cite{Zibaeenejad2019}, the authors investigated how to incrementally take product between an abstraction and multiple DBAs. While this can potentially reduce the overall product system size, it does not directly provide a reduction in the size of the specification. There are other approaches that focus on system decomposition \cite{Kim2018}, hierarchical abstractions \cite{Hsu2018} and parallel computation \cite{Khaled2019tacas}, but not at the specification level. Motivated by this, we pre-process the DBA before applying the proposed control synthesis algorithm. The DBA states that belong to the same strongly connected component (SCC) are treated as a single node. The SCCs of DBA can be computed in linear time w.r.t. the size of the DBA by algorithms such as Tarjan's algorithms \cite{Tarjan1972scc}. States of the DBA are then grouped together to produce a higher-level directed acyclic graph (DAG), in which the nodes present a topological sort. Control synthesis performed in the order of the topological sort reduces the polynomial complexity to linear to avoid unnecessary iterations. Such a process is cost-effective since the size of the DBA is small and a little improvement can result in higher efficiency in control synthesis that involves nonlinear dynamics.

\subsection{Organization and notation}

The rest of the paper is organized as follows. Section \ref{sec:prob} formulates the control synthesis problem. Section \ref{sec:char} provides a characterization of winning sets for DBA specifications. Section \ref{sec:dba} outlines the synthesis procedure and proves theoretical guarantees of soundness and completeness. Section \ref{sec:preprocess} discusses heuristics for pre-processing DBA translations for  controller synthesis. Section \ref{sec:evaluation} presents a number of benchmark examples that demonstrate the performance of the proposed framework. The paper is concluded in Section \ref{sec:conclude}.

\subsubsection*{Notation} Let $\N$, $\Zp$, $\Z^m$, $\mathbb{R}$, and $\mathbb{R}^n$ denote the set of non-negative integers, positive integers, $m$-dimensional integer vectors, real numbers and $n$-dimensional real vectors, respectively; for any set $S$, we denote by $\card{S}$ the cardinality of $S$, $S^*$ and $S^\omega$ the set of all finite and infinite sequences taking values in $S$, respectively; let $\norminf{\cdot}$ be the infinity norm in $\Real^n$ and $\mathcal{B}_\varepsilon(x):=\{y\in \Real^n \sv \norminf{y-x}\leq\varepsilon, x\in\Real^n\}$; given two sets $A,B\subset\Real^n$, $B\setminus A:=\set{x\in B\,\vert\, x\not\in A}$; the Pontryagin difference is defined as $A\ominus B:=\{c\in\Real^n \,\vert\;c+b\in A, \forall b\in B\}$; for a vector $V$ of size $\card{V}$, the $i$th element of $V$ is $V[i]$ ($i\in\set{1,\dots,|V|}$).

\section{Problem Formulation}\label{sec:prob}
\subsection{Linear temporal logic}
Linear temporal logic (LTL) is a formalism defined over a set of \emph{atomic propositions} (AP), which are true or false statements. LTL consists of propositional logic operators, e.g., true ($\top$), false ($\bot$), {\em negation\/} ($\neg$), {\em disjunction\/} ($\vee$), {\em conjunction\/} ($\wedge$), and temporal operators, e.g., {\em next\/} ($\bigcirc$), {\em until\/} ($\until$). Based on these basic operators, {\em eventually} $\eventually=\top\until$ and {\em always} $\always=\neg\eventually\neg$ are frequently used in various scenarios. The semantics of $\ltlx$ is defined on \emph{infinite words} over the alphabet $2^{AP}$, which are infinite sequences of propositions. We refer the reader to \cite[Chapter 5]{baier2008principles} for a complete description of the syntax and semantics of LTL.

\subsection{B\"uchi automata}
\begin{defn}[\cite{Gradel2002}]
  A \textit{B\"uchi automaton} (BA) is a tuple $\A=(Q,\Sigma,r,q_0,F)$, where $Q$ is a finite set of states, $\Sigma$ is a finite alphabet, $r:Q\times\Sigma\to 2^Q$ is the state transition function, $q_0$ is the initial state, and $F\subseteq Q$ is a set of accepting states.
\end{defn}

A \emph{run} of $\A$ is an infinite sequence of states in $Q$ under an input word, denoted by $\varrho=\set{v_i}_{i=0}^\infty$, where $v_i\in Q$ for all $i$. Let $\varrho[i]:=v_i$ and $\varrho[i,j]:=v_i\dots v_j$, $0\leq i<j$. A run $\varrho$ is \emph{successful} for $\A$ if and only if $\varrho$ visits at least one of the states in $F$ infinitely many times, i.e., ${\rm Inf}(\varrho)\cap F\neq\emptyset$, where ${\rm Inf}(\varrho)=\set{v\in Q\sv \forall i, \exists j>i, \text{s.t.}\;v=\varrho[j]}$ represents the set of states occurring infinitely many times during the run $\varrho$. The corresponding input word $\word=\set{\sigma_i}_{i=0}^\infty\in\Sigma^\omega$ is \emph{accepted} by $\A$. If $r(q,\sigma)$ is a singleton or an empty set for all $q\in Q$ and $\sigma\in\Sigma$, then $\A$ is \emph{deterministic} ($\A$ is called a DBA); otherwise $\A$ is \emph{non-deterministic} ($\A$ is called an NBA).

An automaton $\A=(Q,\Sigma,r,q_0,F)$ can be presented as a \emph{directed graph} $\graph=(V,E)$, where $V=Q$ is a set of nodes and $E$ is a set of directed edges. We further define the set of outgoing edges $\outedge(v)=\set{\sigma\in\Sigma\sv r(v,\sigma)\neq\emptyset}$.

Every LTL formula $\varphi$ built on a set of atomic propositions $AP$ has an equivalent B\"uchi automaton (NBA), which accepts the words specified by $\varphi$. For the most part of this paper, we focus on the ones that can be translated to DBA and denote by $\A_\varphi=(Q,\Sigma,r,q_0,F)$ an equivalent DBA of an LTL formula $\varphi$, where $\Sigma=2^{AP}$ is the input alphabet. Such LTL formulas are said to be  \emph{DBA-recognizable}. An input symbol $\sigma\in\Sigma$ is usually represented by a propositional formula over the set $AP$ of atomic propositions for $\varphi$.
Without loss of generality, we assume that $\A_\varphi$ is \emph{total}, i.e., 
$\outedge(q)\neq\emptyset,\,\forall q\in Q$, since we can always construct a total automaton for any BA \cite{baier2008principles}.

\subsection{Discrete-time continuous-state systems}
Consider the discrete-time system 
$$\sys\triangleq\langle\X,\act,\D,R,AP,L\rangle,$$
where $\X\subset\Real^n$ and $\act\subset\Real^m$ are non-empty compact sets of states and control inputs, respectively, $\D=\set{d\in\Real^n\mid \norminf{d}\leq\delta}$
($\delta\geq 0$) is a set of perturbations, $AP$ is a set of atomic propositions, $L:\X\to 2^{AP}$ is a \emph{labeling function}, which associates properties to every state in $\X$, and $R\subseteq \X\times\act\times\X$ is a transition relation such that $(x,u,x')\in R$ if and only if there exists $d\in\D$:
  \begin{align}\label{eq:df}
    x'=f(x, u)+d, \quad x,x'\in\X, u\in\act,
  \end{align}
  where $f:\,\mathbb{R}^n\times\Real^m \to \mathbb{R}^n$.

Assume that the disturbance $d\in\D$ is independent of state $x$ and control input $u$. We call system $\sys$ \emph{deterministic} if set $\D$ is $\set{0}$, denoted by $\sys^0$, otherwise $\sys$ is \emph{non-deterministic}. We denote by $\sys^\delta$ the non-deterministic system with perturbations bounded by amplitude  $\delta$.

A sequence of control inputs $\uu=\set{u_t}_{t=0}^\infty$, where $u_i\in\act$ for all $i\in\N$, is called a \emph{control signal}. Similarly, we denote by $\dd=\set{d_t}_{t=0}^\infty$ a sequence of disturbances. A solution of system $\sys$ is denoted by an infinite sequence of states $\x=\set{x_t}_{t=0}^\infty$, which is generated by a control signal $\uu$, disturbance $\dd$ and an initial condition $x_0\in\X$ according to (\ref{eq:df}). The \emph{trace} of a solution $\x$ is defined as $\trace(\x)=\set{L(x_t)}_{t=0}^\infty$.
A trace of a solution is an infinite word. By evaluating the trace of a system solution, an LTL formula can be verified for a control system $\sys$ or realized by designing a control strategy.

\subsection{LTL/DBA control synthesis problem}

\begin{defn}\label{def:ctlr}
  A \emph{finite-memory control strategy} of system $\sys$ is a partial function 
   \begin{equation} \label{eq:ctlr}
     \kappa:\,\X^*\to 2^{\act}.
   \end{equation}
   A control strategy $\kappa$ is called \emph{memoryless} if it only takes in the current state as the input, i.e.,
\begin{align}\label{eq:localctlr}
  \kappa:\,\X\to 2^{\act}.
\end{align}
\end{defn}

A control signal $\uu=\set{u_t}_{t=0}^\infty$ is said to \emph{conform to} a control strategy $\kappa$ if $u_t\in \kappa(x^*_t)$, $x^*_t=\set{x_i}_{i=0}^t\in\X^*$ for all $t\ge 0$. The infinite sequence $\set{x_i}_{i=0}^\infty$ is the resulting solution of $\sys$.

If there exists an initial condition $x_0\in\X$ and a control strategy $\kappa$ such that, for any control signal that conforms to $\kappa$, the resulting traces of system $\sys$ are guaranteed to satisfy a given LTL formula $\varphi$, we say $\varphi$ is \emph{realizable} for $\sys$. Such a control strategy $\kappa$ \emph{realizes} $\varphi$ for $\sys$. The set of all initial conditions, from which a control strategy $\kappa$ can realize $\varphi$, is called the \emph{winning set} of system $\sys$ w.r.t. $\varphi$, written as $\win_\sys(\varphi)$. If $\win_\sys(\varphi)\neq \emptyset$, then $\varphi$ is realizable for $\sys$.

Our objective is to solve the \textbf{LTL/DBA Control Synthesis Problem}:
Consider system $\sys^0$ and an LTL formula $\varphi$.
\begin{enumerate}[(i)]
\item \label{itm:prob1} Determine whether $\varphi$ is realizable for system $\sys^0$;
\item \label{itm:prob2} Synthesize a control strategy such that the trace of any closed-loop system solution satisfies $\varphi$ if possible.
\end{enumerate}

Although we focus on the deterministic system $\sys^0$, the main result of this paper presented in Section~\ref{sec:dba} is established on the the winning sets of both deterministic and non-deterministic systems.

\section{Characterization of Winning Sets On The Continuous State Space}\label{sec:char}

The objective of this section is to characterize the winning set of system $\sys$ w.r.t. a DBA-recognizable LTL formula over its continuous state space.

\subsection{$\sys$-domains of automaton states}
Let $\A_\varphi$ be an equivalent DBA of an LTL formula $\varphi$ and $q\in Q$ be an arbitrary state of $\A_\varphi$. 
Since $\A_\varphi$ is deterministic and total, every state has at least one outgoing edge and
\begin{equation}\label{eq:outedges}
    \bigvee_{\sigma\in\outedge(q)}\sigma=\top,\;\sigma\wedge\sigma'=\bot,\;\forall\sigma,\sigma'\in\outedge(q)\;\text{s.t.}\;\sigma\neq\sigma'.
\end{equation}

We consider traces of system $\sys$ as input words to $\A_\varphi$. Hence, given a control signal $\uu=\set{u_t}_{t=0}^\infty$ and a sequence of disturbance $\dd=\set{d_t}_{t=0}^\infty$, the resulting run $\varrho=\set{v_t}_{t=0}^\infty$ of $\A_\varphi$ is obtained explicitly by (for all $t\in\Zp$)
\begin{align}\label{eq:qx}
  \begin{cases}
    v_0=q_0,\;v_t= r(v_{t-1},L(x_{t-1})),\; v_t\in Q,\\
    x_t=f(x_{t-1},u_{t-1})+d_{t-1},\; x_t\in\X,
  \end{cases}
\end{align}
where the system and DBA states are updated only at discrete time instances. 
An intuitive illustration of (\ref{eq:qx}) is given in Fig.~\ref{fig:qx}.
\vspace{-5mm}
\tikzset{
  qnode/.style={
    circle,
    draw, thick,
    minimum size=2.5em,
    inner sep=2pt,
    text centered,
    font=\rmfamily,
  },
}
\begin{figure}[ht]
  \centering
  \begin{subfigure}[ht]{\linewidth}
    \centering
    \resizebox{0.4\linewidth}{!}{%
    \begin{tikzpicture}[->,>=stealth',auto,node distance=2cm,semithick]
    \node[state] (v_0) {$q_2$};
    \node[state] (v_1) [right of=v_0] {$q_1$};
    \coordinate[left of=v_0,node distance=1.5cm] (in);
    \coordinate[below right of=v_1,node distance=1.5cm] (out);
    \coordinate[above right of=v_1,node distance=1.5cm] (in1);
    \path (v_0) edge [loop above] node {$b$} (v_0)
    edge node {$c$} (v_1)
    (v_1) edge [loop above] node {$a$} (v_1)
    edge [dashed] node {} (out)
    (in) edge [dashed] node {} (v_0)
    (in1) edge [dashed] node {} (v_1);      
    \end{tikzpicture}}
    \caption{Part of a DBA $\A_\varphi$.}
    \label{fig:qx-dba}
  \end{subfigure}
  \begin{subfigure}[ht]{\linewidth}
    \centering
    \includegraphics[width=0.8\columnwidth]{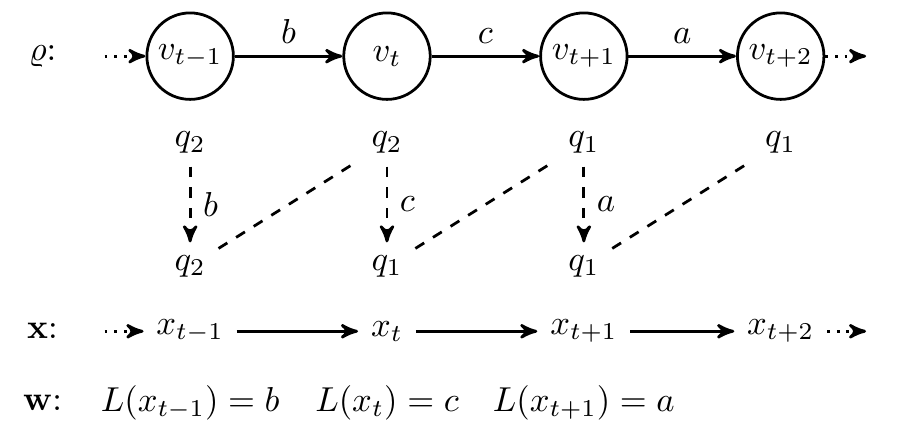}
    \caption{The interpretation of (\ref{eq:qx}).}
    \label{fig:qx-flow}
  \end{subfigure}
  \caption{The connection between system $\sys$ and an equivalent DBA $\A_\varphi$ of a given LTL formula $\varphi$. Assume $v_{t-1}=q_2$ and $L(x_{t-1})=b$. (b) shows how the partial sequence $v_{t-1} v_t v_{t+1} v_{t+2}$ is driven by $x_{t-1} x_t x_{t+1} x_{t+2}$ according to the relevant part of $\A_\varphi$ shown in part (a).}
  \label{fig:qx}
\end{figure}
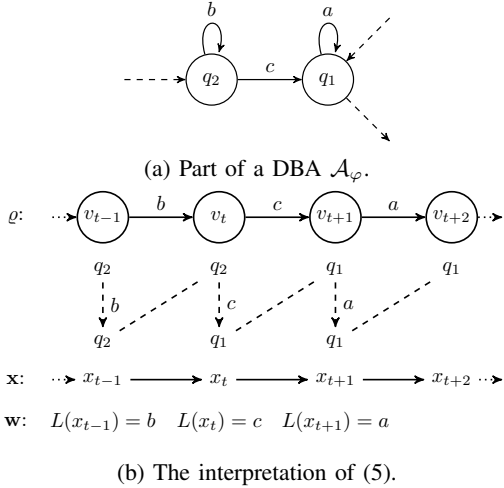

It is easy to see from (\ref{eq:qx}) that the connection between the dynamical system $\sys$ and the targeted DBA $\A_\varphi$ is through the labeling function $L$, which divides the system state space into finitely many areas of interest.
\begin{defn} \label{def:part}
Given a set $\Omega \subseteq\Real^n$ and a positive integer $N$, a finite collection of sets $\P=\{P_1,P_2,\cdots,P_N\}$ is said to be a \emph{partition} of $\Omega$ if (i) $P_i \subseteq \Omega$, for all $i\in\set{1,\cdots,N}$; (ii) $\interior{P_i} \cap \interior{P_j}=\emptyset$ for all $i,j\in\set{1,\cdots,N}$,  where $\interior{P_i}$ denotes the interior of set $P_i$, $i\neq j$; (iii) $\Omega \subseteq \bigcup_{i=1}^N P_i$. Each element $P_i$ of the partition $\P$ is called a \emph{cell}.
\end{defn}

Let $\set{\alpha_1,\cdots,\alpha_N}\in 2^{2^{AP}}$ be such that
\begin{align}\label{eq:Lpart}
  \bigvee_{i=1}^{N}\alpha_i=\top,\quad \alpha_i\wedge\alpha_j=\bot,\,i\neq j.
\end{align}
Then we can obtain a partition $\P_0=\set{P_1,P_2,\cdots,P_{N}}$ of the state space $\X$, where
\begin{align}\label{eq:P0}
  P_i:=L^{-1}(\alpha_i)=\set{x\in\X\sv L(x)=\alpha_i}.
\end{align}
This is to say that all the states inside a cell are assigned the same atomic proposition. Additionally, $L^{-1}(\top)=\X$.

In order to control system $\sys$ so that the resulting traces are accepted by $\A_\varphi$, each transition along the successful runs of $\A_\varphi$ needs to be executed sequentially. Those transitions, however, cannot be assigned deliberately as they have to satisfy the transition relation $R$ of system $\sys$. This also implies that, for each $q\in Q$, the corresponding system state $x$ is restricted to a certain subset of the state space $\X$. To capture such a set, we introduce the following definition.

\begin{defn}\label{def:sdom}
  Let $q\in Q$ and $x\in\X$ be a state of DBA $\A_\varphi$ and system $\sys$, respectively. Then $x$ belongs to the \emph{$\sys$-domain} of $q$, written as $W_\sys(q)$, if and only if there exists a control strategy $\kappa$ in the form of (\ref{eq:ctlr}) such that any run $\varrho=\set{v_t}_{t=0}^\infty$ of $\A_\varphi$ generated by (\ref{eq:qx}) under a control signal conform to $\kappa$ with $v_0=q$ and $x_0=x$ satisfies that ${\rm Inf}(\varrho)\cap F\neq\emptyset$.
\end{defn}

The winning set of an LTL formula $\varphi$ is, by definition, the $\sys$-domain of the initial state $q_0$ of $\A_\varphi$, i.e., $\win_\sys(\varphi)=W_\sys(q_0)$. For any other state $q\in\A_\varphi$, the $\sys$ domain $W_\sys(q)$ is equivalent to the winning set when $q$ is the initial state. Therefore, the problem of computing $\win_\sys(\varphi)$ can be reduced to computing $W_\sys(q_0)$.

\subsection{Characterization of $\sys$-domains}
It is easy to see from (\ref{eq:qx}) that the connection between dynamical system $\sys$ and the targeted DBA $\A_\varphi$ is through the labeling function $L$. Since the outgoing edges satisfy (\ref{eq:outedges}), which is the same as (\ref{eq:Lpart}), the set $\outedge(q)$ of every state $q\in Q$ forms a partition $\P(q)=\set{L^{-1}(\sigma)}_{\sigma\in\outedge(q)}$ of the state space $\X$ through the labeling function $L$ as defined in (\ref{eq:P0}).

Because of the graph structure of a DBA $\A_\varphi$, $\sys$-domains of different automaton states are related with one another by the transitions among them (see Fig.~\ref{fig:qx}). Any state $x\in W_\sys(q)$ can be controlled to the $\sys$-domain of one of the succeeding states of $q$ in $\A_\varphi$. Suppose that the current state of the DBA in Fig.~\ref{fig:qx-dba} is $q_2$ and the system state is $x$. Let $W_\sys(q_2)$ and $W_\sys(q_1)$ be the $\sys$-domains of $q_2$ and $q_1$, respectively. Then $x\in W_\sys(q_0)$ if and only if $\exists u\in\act$ such that,  $\forall d\in\D$, 
\begin{align*}
\begin{cases}
  f(x,u)+d\in W_\sys(q_0),& \text{if}\;L(x)=b,\\
  f(x,u)+d\in W_\sys(q_1),& \text{if}\;L(x)=c.
\end{cases}
\end{align*}

To illustrate this relationship between $\sys$-domains of connected automaton states, we rely on the following definitions.

\begin{defn}\label{def:pre}
  Given a set $X\subset \X$, the \emph{predecessor} of $X$ w.r.t. system $\sys$ is a set of states defined by
  \vspace{-1mm}
  \begin{align*}
    \pre^\delta(X)=\left\{x\in \X \sv \exists u\in\act, \forall d\in\D,f(x,u)+d\in X \right\},
  \end{align*}
   where $\delta\geq 0$ is the bound of disturbances in $\D$.
\end{defn}

For the sake of simplicity, we denote by $\pre$ the predecessor when $\delta=0$, and $\pre^\delta(Y\,|\,X)\triangleq\pre^\delta(Y)\cap X$ for $Y,X\subseteq\X$. 

We also define the set of valid control values that lead to one-step transition to $X$ for an $x\in\pre^\delta(X)$ as
\begin{align*}
  \Pi^\delta_X(x)=\set{u\in\act\sv f(x,u)+d\in X,\forall d\in\D}.
\end{align*}
It is straightforward that $\pre$ is monotonic, i.e., $\pre(A)\subset \pre(B)$ if $A\subset B\subset \Real^n$, and we have in general $\pre(A)\cup\pre(B)\subseteq\pre(A\cup B)$. By definition, $\pre^\delta(X)=\pre(X\ominus\Ball{\delta})$. Such properties  also hold for $\pre^\delta$.

Let $\M$ be an $n_1$ by $n_2$ ($n_1,n_2>0$) matrix of symbols from $\Sigma$, and $V$ and $W$ be two vectors of subsets of $\X$ of length $n_2$. Denote by $m_{ij}$ the element at the $i$th row and $j$th column of $\M$. Define
\begin{align}
  W+V&\triangleq \begin{bmatrix}W[1]\cup V[1]& \cdots & W[n_2]\cup V[n_2]\end{bmatrix}^T,\label{eq:w+v}\\
  W-V&\triangleq \begin{bmatrix}W[1]\setminus V[1] & \cdots & W[n_2]\setminus V[n_2]\end{bmatrix}^T, \label{eq:w-v}\\
  V\preceq W&\triangleq V[i]\subseteq W[i],\;i=1,\dots,n_2, \label{eq:v<w}\\
  W=V&\triangleq W[i]=V[i],\;i=1,\dots,n_2,\label{eq:w=v}
\end{align}
\begin{align}
  \label{eq:T}
  \T^\delta(\M,W)=W',
\end{align}
where
\begin{align*}
  W'[i]=\bigcup_{j=1}^{n_2}\pre^\delta\left(W[j]\,|\,L^{-1}(m_{ij})\right),\; i=1,\dots,n_1.
\end{align*}

For the nominal system $\sys$,  where $\delta=0$, we use $\T$ for the sake of simplicity. Based on the properties of predecessor maps, the operator $\T^\delta$ satisfies the following properties.
\begin{prop}\label{prop:Tproperties}
  Given a matrix $\M$ of symbols and vectors $V,W$ of subsets of $\X$ that match in dimension for operator $\T^\delta$ defined in (\ref{eq:T}) with $\delta\geq 0$ and $0\leq \delta_1\leq\delta_2$,
  \begin{enumerate}[(i)]
  \item \label{itm:T1}$\T^\delta(\M,V)\preceq \T^\delta(\M,W)$ if $V\preceq W$,
  \item \label{itm:T3}$\T^{\delta_2}(\M,W)\preceq \T^{\delta_1}(\M,W)\preceq \T(\M,W)$.
  \end{enumerate}
\end{prop}

\begin{IEEEproof}
  To show (\ref{itm:T1}), assume that $\M$ is of size $n_1\times n_2$ and $W$, $V$ of size $n_2\times 1$. As defined in (\ref{eq:v<w}), $V\preceq W$ means $V[j]\subseteq W[j]$. By the monotonicity of $\pre^\delta$, we have $\bigcup_{i=1}^{n_1}\pre^\delta(V[j]\,|\,L^{-1}(m_{ij}))\subseteq \bigcup_{i=1}^{n_1}\pre^\delta(W[j]\,|\,L^{-1}(m_{ij}))$, which gives (\ref{itm:T1}).

  Property (\ref{itm:T3}) is straightforward by the fact that $\pre^\delta(A\ominus\Ball{\delta_2})\subseteq\pre^\delta(A\ominus\Ball{\delta_1})\subseteq\pre^\delta(A)$ for all $A\subseteq\X$.
\end{IEEEproof}

The graph representation of a DBA can be coded into a matrix of symbols, which is given in the following definition.
\begin{defn}\label{def:M}
  Given a DBA $\A_\varphi$ with the set of indexed states $Q$ with $Q[i]$ being the $i$th element ($i\in\set{1,\cdots,\card{Q}}$), the \emph{transition matrix} $\M_\varphi$ of $\A_\varphi$ is a $\card{Q}$ by $\card{Q}$ matrix of symbols from $\Sigma$. The element $m_{ij}$ in the $i$th row and $j$th column ($i,j=1,\dots,\card{Q}$) of $\M_\varphi$ is given by
\begin{align}\label{eq:M}
  m_{ij}=
  \begin{cases}
    \sigma, & Q[j]=r(Q[i],\sigma),\sigma\in \Sigma,\\
    e, & \text{otherwise},
  \end{cases}
\end{align}
where $e\in\Sigma$ denotes an empty symbol with $L^{-1}(e)=\emptyset$.
\end{defn}
As defined in (\ref{eq:T}), the operator $\T^\delta$ computes predecessors according to the transition relation coded in $\M_\varphi$.

To track the control values that can activate the transitions, we also define a vector $\K=\begin{bmatrix}\kappa_1 & \dots & \kappa_{n_1}\end{bmatrix}$ of memoryless control strategies, where ($i=1,\dots,n_1$)
\begin{align}
  \kappa_i(x)=\bigcup_{j=1}^{n_2}\Pi_{W[j]}^{\delta}(x),\; \forall x\in W'[i].\label{eq:kappapre}
\end{align}

For a DBA $\A_\varphi$, the dependencies among the $\sys$-domains can be captured by using the operator $\T^\delta$ and the transition matrix $\M_\varphi$. Suppose $V$ is a vector of subsets of states in the state space $\X$, where each element $V[i]$ of $V$ is a goal region for $Q[i]$. Then $\T^\delta(\M_\varphi,V)$ by definition is the vector of sets that can be controlled into $V$ for system $\sys^\delta$ in one step.

\begin{prop}\label{prop:sdomfp}
  Let $\M_\varphi$ be the transition matrix of a DBA $\A_\varphi$ and $\W_\sys$ be the vector of $\sys$-domains of $\A_\varphi$ for $\sys^\delta$. Then $\W_\sys=\T^\delta(\M_\varphi,\W_\sys)$.
\end{prop}
\begin{IEEEproof}
  Let $V=\T^\delta(\M_\varphi,\W_\sys)$ and $i\in\set{1,\dots,\card{Q}}$.
  We first show that $V\preceq\W_\sys$. By the definition of $\T^\delta$, any state $x\in V[i]$
  can be controlled into $\bigcup_{j=1}^{\card{Q}} W_\sys(Q[j])$ after one step by some $u\in\act$ for all $d\in\D$. This implies that for any state $x\in W_\sys(q)$ (any $q\in Q$), there exists a run $\varrho$ with $\varrho[t]=q$ at some $t\in\Zp$, which is generated according to (\ref{eq:qx}), such that $\varrho$ visits $F$ infinitely often. Hence, $x\in W_\sys(Q[i])$ by Definition \ref{def:sdom}, and $V[i]\subseteq W_\sys(Q[i])$. Since $i$ is arbitrary, $V\preceq\W_\sys$.
  
  Next we show that $\W_\sys\preceq V$. Suppose that there is an $x\in W_\sys(Q[i])$ but $x\notin V[i]=\bigcup_{j=1}^{\card{Q}}\pre^\delta\left(W_\sys(Q[j])\mid L^{-1}(m_{ij})\right)$. Then by Definition \ref{def:pre}, for all $j\in\set{1,\dots,\card{Q}}$ and $u\in\act$, there exists $d_{j,u}\in\D$ such that $x'=f(x,u)+d_{j,u}\notin W_\sys(Q[j])$ or $x\notin L^{-1}(m_{ij})$. If $x\notin L^{-1}(m_{ij})$, then the transition from $Q[i]$ to $Q[j]$ will not happen. If $x'=f(x,u)+d_{j,u}\notin W_\sys(Q[j])$, then by Definition \ref{def:sdom} there exists no control strategy that the rest of the run with system state starting from $x'$ will visit $F$ infinitely often.
  Hence, $x\notin W_\sys(Q[i])$, which contradicts the assumption. Therefore, by (\ref{eq:v<w}) and (\ref{eq:w=v}), $V\preceq\W_\sys$ and $\W_\sys\preceq V$ gives $\W_\sys=V$.
\end{IEEEproof}

\begin{rem}
  Proposition \ref{prop:sdomfp} is a necessary condition for a vector $W$ to be $\W_\sys$, and $\W_\sys$ may not be the unique fixed point of $\T^\delta(\M_\varphi,\cdot)$ w.r.t. a transition matrix $\M_\varphi$. By Definition \ref{def:sdom}, $\W_\sys$ is the maximal fixed point of $\T^\delta(\M_\varphi,\cdot)$ that satisfies the B\"uchi condition.
\end{rem}

We now illustrate previous sections with an example.
\begin{exmp}\label{exp:running}
   Consider a system $\sys$ with $\X=[0,2]$, $\act=[-0.9, -0.8]$, $\delta=0.01$, $AP=\set{a_1, a_2}$, and $L(a_1)=[0.1, 0.2]$, $L(a_2)=[0.5, 0.6]$ and $R$ determined by 
   \begin{align*}
       x'=u(x-1)+1+d,\quad x,x'\in\X, u\in\act, d\in\D.
   \end{align*}
   Let the specification be $\varphi=\eventually(a_1\wedge\eventually a_2)$, which can be translated into a DBA shown in Fig.~\ref{fig:smallexp}.
   \begin{figure}[ht]
      \centering
      \begin{minipage}{0.5\columnwidth}
        \resizebox{1\linewidth}{!}{%
        \begin{tikzpicture}[->,>=stealth',auto,node distance=2cm,semithick,baseline=-30pt]
          \node[state,initial] (v_2) {$q_2$};
          \node[state] (v_1) [right of=v_2] {$q_1$};
          \node[state,accepting] (v_0) [right of=v_1] {$q_0$};
          \path (v_0) edge [loop above] node {$\top$} (v_0)
          (v_1) edge [loop above] node {$\neg a_2$} (v_1)
          edge node {$a_2$} (v_0)
          (v_2) edge [loop above] node {$\neg a_1$} (v_2)
          edge node {$a_1$} (v_1);
        \end{tikzpicture}}
      \end{minipage}%
      \resizebox{0.5\linewidth}{!}{%
      \begin{minipage}{0.5\columnwidth}
      \begin{align*}
        \M_{\varphi}=
        \begin{bmatrix}
          \neg a_1 & a_1 & e\\
          e & \neg a_2 & a_2\\
          e & e & \top\\
        \end{bmatrix}
      \end{align*}
      \end{minipage}}
      \caption{The DBA of $\varphi$ and $\M_{\varphi}$ with the order $q_2,q_1,q_0$.\vspace{-3mm}}
      \label{fig:smallexp}
    \end{figure}
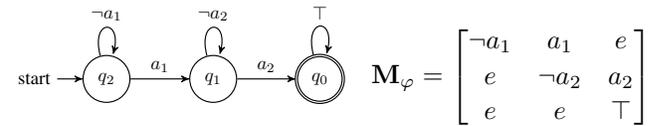
    
    By Definition \ref{def:sdom}, $W_\sys(q_0)=\X$, $W_\sys(q_1)$ and $W_\sys(q)$ are the sets of states that can reach $L(a_2)$ and $L(a_1)\cap W_\sys(q_1)$, respectively, and they can be computed exactly: $W_\sys^0(q_1)=[0, 0.6]\cup[1.444, 2]$, $W_\sys^0(q_2)=[0, 0.012]\cup[0.1,0.2]\cup[1.889, 2]$, $W_\sys^\delta(q_1)=[0, 0.483]\cup[0.5,0.6]\cup[1.456, 2]$, and $W_\sys^\delta(q_2)=[0.1,0.2]\cup[1.9, 2]$. One can verify Proposition \ref{prop:sdomfp}:
    {
    \begin{align*}
        W_\sys(q_1)=\pre(W_\sys(q_0)|L(a_2))\cup\pre(W_\sys(q_1)|L(\neg a_2)),\\
        W_\sys(q_2)=\pre(W_\sys(q_1)|L(a_1))\cup\pre(W_\sys(q_2)|L(\neg a_1)).
    \end{align*}}
\end{exmp}

\section{DBA Control Synthesis Is Robustly Complete}\label{sec:dba}

The determination of $\sys$-domains for system $\sys^0$ involves the computation of the operator $\T$, which is nontrivial under general nonlinear dynamics. Even if $\T(\M_\varphi, V)$ can be computed precisely for a vector $V$, the computation of finding a fixed point is usually iterative and not guaranteed to terminate in a finite number of iterations, because system $\sys^0$ contains infinite number of states. Hence, solving the original DBA control synthesis problem is practically difficult.

In this section, we relax the original control synthesis problem based on the following definition.
\begin{defn}
  An LTL formula $\varphi$ is said to be \emph{$\delta$-robustly realizable} for system $\sys^0$ if it is realizable for $\sys^\delta$. If $\delta>0$, then $\varphi$ is called \emph{robustly realizable} for $\sys^0$.
\end{defn}

\textbf{Relaxed LTL/DBA Control Synthesis Problem}:
Consider system $\sys^0$ and an LTL formula $\varphi$. Answer one of the two following questions:
\begin{enumerate}[(i)]
\item Find a control strategy if $\varphi$ is robustly realizable for $\sys^0$.
\item Verify that $\varphi$ is not realizable for $\sys^\delta$ with $\delta>0$.
\end{enumerate}

We will show later that control synthesis w.r.t. a DBA-recognizable LTL formula $\varphi$ can be made sound and \emph{robustly complete} in the sense of solving the relaxed problem.

\subsection{Approximation of $\T$ via interval computation}
Predecessors can be inner-approximated within a given precision by using the following Algorithm~\ref{alg:cpre} (originally given in \cite{Li2018rocs}). A branch-and-bound scheme is used: the intervals are bisected to left ($L[x]$) and right ($R[x]$) parts only when they are undetermined to be within the predecessor (line 12-13). The parameter $\varepsilon$ is used to control the minimal width of an interval. Hence, such interval-based computation automatically divides the state space into a finite number of intervals, which form a partition of the state space. The output is an inner-approximation of $\pre(Y)\cap X$ (for any $X,Y\subseteq\X$) represented by a finite set of intervals, denoted by $[\underline{\pre}_\mu]^\varepsilon(Y|X)$. Particularly, $[\underline{\pre}_\mu]^\varepsilon(Y)=[\underline{\pre}_\mu]^\varepsilon(Y|\X)$.
\begin{algorithm}[htbp]
  \caption{$\underline{X}=[\underline{\pre}_\mu]^\varepsilon(Y|X)$}
  \label{alg:cpre}
  \begin{algorithmic}[1]
    \State $\underline{X}\leftarrow\emptyset,\Delta X\leftarrow\emptyset, X_c\leftarrow\emptyset, List\leftarrow X$
    \While{$List\neq \emptyset$}
    \State $[x]\leftarrow List.first$
    \If{$[f]([x],u)\cap Y=\emptyset$ for all $u\in[\act]_{\mu}$}
    \State $X_c\leftarrow X_c\cup [x]$
    \ElsIf{$[f]([x],u) \subseteq Y$ for some $u\in[\act]_{\mu}$}
    \State $\underline{X}\leftarrow \underline{X}\cup [x]$
    \Else
    \If{$\text{wid}([x])<\varepsilon$}
    \State $\Delta X\leftarrow \Delta X\cup [x]$
    \Else
    \State $\{L[x],R[x] \}=Bisect([x])$
    \State $List.add(\set{L[x],R[x]})$
    \EndIf
    \EndIf
    \EndWhile
  \end{algorithmic}
\end{algorithm}

In Algorithm~\ref{alg:cpre}, $[f]([x],u)$ computes an interval over-approximation of the set $f([x],u)$, where $[x]$ denotes an interval (See \cite{Li2018rocs} for details on related interval computation and further explanation of Algorithm~\ref{alg:cpre}). The infinite set of control inputs is under-sampled by a grid size $\mu$:
\begin{align}
  \label{eq:cset}
  [\act]_{\mu}\triangleq\mu\Z^m\cap\act,\; \mu\Z^m=\set{\mu z\sv z\in \Z^m, \mu>0}
\end{align}

It has been shown that $[\underline{\pre}_\mu]^\varepsilon(Y|X)$ is lower bounded by using a set of under-sampled control values (\ref{eq:cset}) if the following assumption is satisfied.

\begin{assum}\label{asp:lipschitz}
  There exists a Lipschitz constant $\rho>0$ for the function $f:\Real^n\times\Real^m\to\Real^n$ in (\ref{eq:df}) such that for all $u,v\in\act$ and $x,y\in\X$: $\norminf{f(x,u)-f(y,v)}\leq \rho(\norminf{x-y}+\norminf{u-v})$.
\end{assum}

\begin{rem}\label{rem:rho}
If $f$ is differentiable in both arguments, the constant $\rho$ can be estimated as
{ $$\max\set{\sup_{u\in\act}\sup_{x\in\X}\norm{J_x f(x,u)}, \sup_{x\in\X}\sup_{u\in\act}\norm{J_u f(x,u)}},$$}
where $J_x f(x,u)$ ($J_u f(x,u)$) is the Jacobian of $f$ w.r.t. $x$ ($u$) for a fixed $u$ ($x$).
\end{rem}

\begin{lem}[\cite{Li2020Reachstay}]\label{lem:intervalpre-ucont1}
  Consider system $\sys$. Let $\mu$ be a parameter given in (\ref{eq:cset}) and $X,Y\subseteq\X$ be compact. If Assumption \ref{asp:lipschitz} holds, then
   \begin{align*}
     \pre(Y\ominus\Ball{\rho(\varepsilon+\mu)}|X)\subseteq [\underline{\pre_\mu}]^\varepsilon(Y|X)\subseteq \pre(Y|X).
   \end{align*}
\end{lem}

An interpretation of Lemma \ref{lem:intervalpre-ucont1} is that the interval-based approximation of a predecessor for system $\sys^0$ is still lower bounded by the one of a perturbed system $\sys^\delta$ with $\delta\geq\rho(\varepsilon+\mu)$, even though the computation only utilizes sampled control inputs and discritized states.
A similar result can be drawn for the interval operator $[\T_\mu]^\varepsilon$ of $\T$, which is defined by replacing $\pre$ with $[\underline{\pre}_\mu]^\varepsilon$.

\begin{lem}\label{lem:intervalT}
  Consider system $\sys$ with an under-sampled control set defined in (\ref{eq:cset}) with the parameter $\mu$. Let $\M$ be a matrix of symbols from $\Sigma$ and $W$ be a vector of subsets of $\X$, and $\M$ and $W$ match in dimension. If Assumption \ref{asp:lipschitz} holds, then 
  \begin{align*}
    \T^{\rho(\varepsilon+\mu)}(\M,W)\preceq [\T_\mu]^\varepsilon(\M,W)\preceq \T(\M,W).
  \end{align*}
\end{lem}
\begin{IEEEproof}
  Let $V=\T(\M,W)$, $V'=[\T_\mu]^\varepsilon(\M,W)$ and $V''=\T^{\rho(\varepsilon+\mu)}(\M,W)$. Assume that $\M$ and $W$ are $n_1\times n_2$ and $n_2\times 1$, respectively. Then for all $i\in\set{1,\cdots, n_1}$, $V[i]=\bigcup_{j=1}^{n_2}\pre\left(W[j]|L^{-1}(m_{ij})\right)$, $V'[i]=\bigcup_{j=1}^{n_2}[\underline{\pre_\mu}]^\varepsilon\left(W[j]|L^{-1}(m_{ij})\right)$, and $V''[i]=\bigcup_{j=1}^{n_2}\pre^{\rho(\varepsilon+\mu)}\left(W[j]|L^{-1}(m_{ij})\right)$. Lemma \ref{lem:intervalpre-ucont1} gives
  \begin{align*}
    \pre^{\rho(\varepsilon+\mu)}\left(W[j]\right)&=\pre\left(W[j]\ominus\Ball{\rho(\varepsilon+\mu)}\right)\\
                                                 &\subseteq [\underline{\pre_\mu}]^\varepsilon\left(W[j]\right)\subseteq \pre\left(W[j]\right)
  \end{align*}
  for all $j\in\set{1,\cdots,n_2}$. 
  Then we have $V''[i]\subseteq V'[i]\subseteq V[i]$, which shows that $V''\preceq V'\preceq V$ by (\ref{eq:v<w}).
\end{IEEEproof}

\subsection{Robustly complete control synthesis}
We now present Algorithm~\ref{alg:buchigame} to practically inner approximate the vector of winning sets $\W_\sys(\varphi)$ of system $\sys^0$ w.r.t. a DBA-recognizable LTL formula.

\begin{algorithm}[htbp]
  \caption{$W,\K=\textsc{Sdom}(\M_\varphi,[\T_\mu]^\varepsilon)$}
  \label{alg:buchigame}
  \begin{algorithmic}[1]
    \State $n_1=\card{Q}-\card{F}$, $n_2=\card{F}$
    \State $\M_\varphi=\begin{bmatrix}\M_1 \\ \M_2\end{bmatrix}$, $\M_1$ ($n_1\times (n_1+n_2)$), $\M_2$ ($n_2\times (n_1+n_2)$)
    \State $\widetilde{Y}(n_1\times 1)$, $\K(1\times\card{Q})$
    \State $\widetilde{Z}(n_2\times 1)$, $\widetilde{Z}[j]\leftarrow\X$, $j\in\set{1,\cdots,n_2}$
    \Repeat
    \State $Z\leftarrow\widetilde{Z}$
    \State $\widetilde{Y}[i]\leftarrow\emptyset$, $i\in\set{1,\dots,n_1}$
    \Repeat
    \State $Y\leftarrow\widetilde{Y}$
    \State $\widetilde{Y}\leftarrow Y+[\T_\mu]^\varepsilon(\M_1,\begin{bmatrix}Y\\Z\end{bmatrix})$ \algorithmiccomment{(\ref{eq:w+v}) (\ref{eq:T})}
    \State assign $\K[i](x)$ by (\ref{eq:kappapre}) for all $x\in\widetilde{Y}[i]\setminus Y[i]$ and $i\in\set{1,\cdots,n_1}$
    \Until{$Y=\widetilde{Y}$}
    \State $\widetilde{Z}\leftarrow [\T_\mu]^\varepsilon(\M_2,\begin{bmatrix}Y\\Z\end{bmatrix})$ \algorithmiccomment{(\ref{eq:T})}
    \State assign $\K[n_1+j](x)$ by (\ref{eq:kappapre}) for all $x\in\widetilde{Z}[j]$ and $j\in\set{1,\cdots,n_2}$
    \Until{$Z=\widetilde{Z}$} \algorithmiccomment{(\ref{eq:w=v})}
    \State \textbf{Return} $W\leftarrow\begin{bmatrix}Y\\Z\end{bmatrix}$, $\K$
  \end{algorithmic}
\end{algorithm}

The input arguments of Algorithm~\ref{alg:buchigame} are the transition matrix $\M_\varphi$ of $\A_\varphi$ that reflects the transition relation of system $\sys$ and an operator $[\T_\mu]^\varepsilon$. We assume that the nodes of $\A_\varphi$ are sorted so that accepting nodes rank after nonaccepting ones, and the transition matrix $\M_\varphi$ is divided into 2 matrix blocks $\M_1$, $\M_2$ which represent the transitions from nonaccepting and accepting nodes, respectively. Let $\bar{F}=Q\setminus F$.

Let $l$ and $\nu$ denote the indices of the inner loop and outer loop of Algorithm~\ref{alg:buchigame}, respectively.
$\set{Y_\nu^l}_{l=0}^\infty$ denotes the sequence of vectors generated by the inner loop of the $\nu^{th}$ outer loop, and $\set{Z_\nu}_{\nu=0}^\infty$ denotes the sequence of vectors generated by the  outer loop. It is clear that $Y_\nu^0=\emptyset$ and $Y_\nu^l\preceq Y_\nu^{l+1}$ by (\ref{eq:w+v}) for all $l\in\N$.
It follows that $\set{Y_\nu^l}_{l=0}^\infty$ is increasing for all $\nu\in\N$.
Let $Y_\nu$ be the fixed point of the $\nu^{th}$ outer loop, i.e., $Y_\nu=\bigcup_{l=0}^\infty Y_\nu^l$.

Intuitively, by setting $Y_\nu^0=\emptyset$ for all $\nu$, the inner loop generates the set of states for each nonaccepting node that can be controlled to reach one of the elements in $Z_\nu$.
The outer loop keeps the states in each element of $Z_\nu$ that can still be controlled to reach any of the elements in $Z_\nu$.
The sequences $\set{Y_\nu}_{\nu=0}^\infty$ and $\set{Z_\nu}_{\nu=0}^\infty$ approach the $\sys$-domains of the nonaccepting and accepting nodes, respectively. Define
\begin{align}\label{eq:W}
  W=\begin{bmatrix}Y \\ Z\end{bmatrix}=\bigcap_{\nu=0}^\infty\begin{bmatrix}Y_\nu \\ Z_\nu\end{bmatrix}=\bigcap_{\nu=0}^\infty W_\nu,\quad W_\nu=\begin{bmatrix}Y_\nu\\Z_\nu\end{bmatrix}.
\end{align}

Based on Lemma \ref{lem:intervalT}, we show in Theorem~\ref{thm:maindba} that, by using sufficiently small precision parameter $\varepsilon$ and sampling grid size $\mu$ for the control set in Algorithm~\ref{alg:buchigame}, control synthesis for system $\sys$ w.r.t. DBA-recognizable LTL formulas can be made sound and robustly complete. 
The proof follows the above notation for the sequences generated in Algorithm~\ref{alg:buchigame}.

\begin{thm}\label{thm:maindba}
  Consider system $\sys$ and a DBA $\A_\varphi$. Denote by $\M_\varphi$ the transition matrix of $\A_\varphi$.
  Let $[\T_\mu]^\varepsilon$ be an interval implementation of $\T$ for system $\sys$, where $\varepsilon$ is the precision and $\mu$ is the sampling parameter in (\ref{eq:cset}). Suppose that Assumption \ref{asp:lipschitz} holds. Then Algorithm~\ref{alg:buchigame} terminates in a finite number of iterations.
  Furthermore, if $\rho(\varepsilon+\mu)\leq\delta$, then
  \begin{align}\label{eq:maindba}
    \win_\sys^\delta(\varphi)\subseteq W(q_0)\subseteq\win_\sys(\varphi),
  \end{align}
  where $W(q_0)$ is the element of $W$ corresponding to the initial state $q_0$ of $\A_\varphi$.
\end{thm}
\begin{IEEEproof}
  Let $\W_\sys$ and $\W_\sys^\delta$ be the $\sys$-domains of $\sys^0$ and $\sys^\delta$, respectively. Let $i,i'\in\set{1,\cdots,n_1}$ and $j,j'\in\set{1,\cdots,n_2}$ denote the indices of the elements in $\bar{F}$ and $F$, respectively. We assume $i,i',j,j'$ to be arbitrary throughout the proof.

  \textbf{Claim}: For all $\nu\in \Zp$, $Y_{\nu}\preceq Y_{\nu-1}$ if $Z_\nu\preceq Z_{\nu-1}$. By Proposition~\ref{prop:Tproperties} (\ref{itm:T1}), $Y_{\nu}^0=Y_{\nu-1}^0=\emptyset$, and $Z_\nu\preceq Z_{\nu-1}$, we have
  \begin{align*}
      Y_{\nu}^1&=Y_{\nu}^0+[\T_\mu]^\varepsilon\left(\M_1,\begin{bmatrix}Y_{\nu}^0\\Z_{\nu}\end{bmatrix}\right)\\
      &\preceq Y_{\nu-1}^0+[\T_\mu]^\varepsilon\left(\M_1,\begin{bmatrix}Y_{\nu-1}^0\\Z_{\nu-1}\end{bmatrix}\right)=Y_{\nu-1}^1.
  \end{align*}
  This gives $Y_{\nu}^l\preceq Y_{\nu-1}^l$ for all $l\in\N$. It follows that $Y_{\nu}=\bigcup_{l=0}^\infty Y_{\nu}^l\preceq \bigcup_{l=0}^\infty Y_{\nu-1}^l=Y_{\nu-1}$. Hence, the claim is proved.
  
  To see the finite termination of Algorithm~\ref{alg:buchigame}, we first show that $\set{Z_\nu}_{\nu=0}^\infty$ and $\set{Y_\nu}_{\nu=0}^\infty$ are decreasing by induction. We have $Z_1\preceq Z_0$ since $Z_0[j]=\X$. Then $Y_1\preceq Y_0$ by the claim. Assume $Z_{\nu}\preceq Z_{\nu-1}$ and $Y_{\nu}\preceq Y_{\nu-1}$ for some $\nu\in \Zp$. Then $Z_{\nu+1}=[\T_\mu]^\varepsilon\left(\M_2,W_\nu\right)\preceq [\T_\mu]^\varepsilon\left(\M_2,W_{\nu-1}\right)=Z_\nu$ by Proposition~\ref{prop:Tproperties} (\ref{itm:T1}), and $Y_{\nu+1}\preceq Y_{\nu}$ by the claim, which shows that $\set{Z_\nu}_{\nu=0}^\infty$ and $\set{Y_\nu}_{\nu=0}^\infty$ are decreasing.
  
  Since the widths of the intervals that partition the state space are lower bounded by $\varepsilon$, each element in $Y$ and $Z$ contains finitely many intervals. Hence, Algorithm~\ref{alg:buchigame} will terminate in a finite number of steps because $\set{Y_\nu^l}_{l=0}^\infty$ and $\set{Z_\nu}_{\nu=0}^\infty$ are decreasing. This implies that $\exists N_Z\in\N$ such that $Z_{N_Z}=Z_{N_Z+1}$, and for all $\nu$, $\exists N_Y(\nu)\in\N$ such that $Y_\nu^{N_Y(\nu)}=Y_\nu^{N_Y(\nu)+1}=Y_\nu$ and $Y_{N_Z}=Y_{N_Z+1}$. It follows that
  \begin{align}
    Y_\nu&=Y_\nu+[\T_\mu]^\varepsilon\left(\M_1,W_\nu\right)\; \forall \nu\in\N,\label{eq:yfp}\\
    Z&=[\T_\mu]^\varepsilon\left(\M_2,W\right),\label{eq:zfp}
  \end{align}
  where $Z=Z_{N_Z}$, $Y=Y_{N_Z}^{N_Y}$, and $W=\begin{bmatrix}Y&Z\end{bmatrix}^T$.
  
  By Proposition \ref{prop:Tproperties} (\ref{itm:T3}), if $\rho(\varepsilon+\mu)\leq\delta$, then $\T^\delta(\M,V)\preceq \T^{\rho(\varepsilon+\mu)}(\M,V)$ for any consistent vector $V$ of sets. Together with Lemma \ref{lem:intervalT}, we have
  \begin{align}
    \T^\delta(\M,V)\preceq[\T_\mu]^\varepsilon(\M,V)\preceq \T(\M,V).\label{eq:key}
  \end{align}
  Let $\varrho$ be a run of $\A_\varphi$ resulting from a solution $\set{x_t}_{t=0}^\infty$ and $\varrho[t]$ denote the DBA state at time $t$.
  
  We now show the soundness, i.e., $W\preceq\W_\sys$. Let $A_\nu^0=Y_\nu^0=\emptyset$ and $ A_\nu^l=A_\nu^{l-1}+\T(\M_1,\begin{bmatrix}A_\nu^l&Z_\nu\end{bmatrix}^T)$ for $l\in\Zp$. By the definition of $\T$, if $\varrho[t]=\bar{F}[i]$, any $x_t\in A_\nu^l[i]$ for $\sys^0$ can trigger the transition in $\A_\varphi$ to $F[j]$ within $l$ steps. 
  By (\ref{eq:key}), $Y_\nu^l\preceq A_\nu^l$ for all $l\in\N$, which leads to $Y\preceq A_{N_Z}^{N_Y(N_Z)}$. That is to say, for $\sys^0$, $\forall x_t\in Y[i]$, $\exists\set{u_k}_{k=t}^{t+N_Y(N_Z)-1}$ such that $\varrho[t']\in F$ for some $t'\in[t, t+N_Y(N_Z)]$. 
  Based on (\ref{eq:zfp}) and (\ref{eq:key}), we have $Z\preceq \T(\M_2,W)$. This means if $\varrho[t]=F[j]$ and $x_t\in Z[j]$ for $\sys^0$, then either (i) $x_{t+1}\in Y[i]$ if $\varrho[t+1]=\bar{F}[i]$, and $F$ can be visited again from $Y[i]$, or (ii) $x_{t+1}\in Z[i']$ if $\varrho[t+1]= F[j']$. Hence, we can conclude that $W\preceq\W_\sys$.
  
  To see $\W_\sys^\delta\preceq W$, we aim to show $x\notin Y[i]\Rightarrow x\notin\W_\sys^\delta(\bar{F}[i])$ and $x\notin Z[j]\Rightarrow x\notin\W_\sys^\delta(F[j])$. Consider arbitrary $t\in\N$ and arbitrary $\nu\in\set{1,\cdots,N_Z}$. We first discuss two situations based on (\ref{eq:key}) and the definition of $\T^\delta$ in (\ref{eq:T}):
  
  (i) $\varrho[t]$ is accepting, i.e., $\varrho[t]=\bar{F}[i]$: if $x_t\notin Y_\nu[i]$, then $\forall u_t\in\act$, $\exists d_t$ with $\norminf{d_t}\leq\delta$ such that
  \begin{align*}
      \begin{cases}
        x_{t+1}\notin Y_{\nu}[i'], & {\rm if}\;\varrho[t+1]=r(\bar{F}[i],L(x_t))\notin F,\\
        x_{t+1}\notin Z_{\nu}[j], & {\rm if}\;\varrho[t+1]=r(\bar{F}[i],L(x_t))=F[j].
      \end{cases}
  \end{align*}
   In addition, by (\ref{eq:yfp}), if $x_t\notin Y_\nu[i]$ and $\varrho[t+1]\notin F$, then $x_t\notin [\T_\mu]^\varepsilon\left(\M_1,W_\nu\right)$. Otherwise $Y_\nu\neq Y_\nu+[\T_\mu]^\varepsilon\left(\M_1,W_\nu\right)$. This implies $\forall u_t\in\act$, $\exists d_t$ with $\norminf{d_t}\leq\delta$ such that $x_{t+1}\notin Y_\nu$ if $\varrho[t+1]\notin F$.
  
  (ii) $\varrho[t]$ is nonaccepting, i.e., $\varrho[t]=F[j]$: if $x_t\notin Z_\nu[j]$, then $\forall u_t\in\act$, $\exists d_t$ with $\norminf{d_t}\leq\delta$ such that
  \begin{align*}
      \begin{cases}
        x_{t+1}\notin Y_{\nu-1}[i], & {\rm if}\;\varrho[t+1]=r(F[j],L(x_t))\notin F,\\
        x_{t+1}\notin Z_{\nu-1}[j'], & {\rm if}\;\varrho[t+1]=r(F[j],L(x_t))=F[j'].
      \end{cases}
  \end{align*}
  If $x_0=x\notin Y[i]$ (i.e., $x_0\notin Y_{N_Z}[i]$), then $x_1\notin Y_{N_Z}[i']$ if $\rho[1]\notin F$ and $x_1\notin Z_{N_Z}[j]$ if $\rho[1]\in F$ by (i). If $\rho[k]= F[j]$ for some $k_0\in\Zp$, then $x_{k_0}\notin Z_{N_Z}[j]$; otherwise $\rho[t]\notin F$ for all $t$, which implies $x\notin\W_\sys^\delta(\bar{F}[i])$. Considering (ii) for $x_{k_0}\notin Z_{N_Z}[j]$, we have $x_{k_0+1}\notin Y_{N_Z-1}[i']$ if $\varrho[k_0+1]=\bar{F}[i']$ and $x_{k_0+1}\notin Z_{N_Z-1}[j']$ if $\varrho[k_0+1]=F[j']$.
  Combining (i) and (ii) in this manner, $\nu$ decreases from $N_Z$ by 1 every time $\varrho$ visits $F$ 
  until $\nu=0$ at some $k_{N_Z}\in\Zp$, and $x_{k_{N_Z}}\notin Y_0[i']$. If $\varrho[k_{N_Z}+1]=F[j]$ for some $j\in\set{1,\cdots,n_2}$, then $x_{k+1}\notin Z_0[j]$, which means is impossible since $Z_0$ is a vector of the full state space of the system. Hence, we have $\varrho[t']\notin F$ for all $t'\geq k_{N_Z}$, which gives $x\notin\W_\sys^\delta(\bar{F}[i])$. The same argument applies to the case in which $x_0=x\notin Z_\nu[j]$ and proves $x\notin\W_\sys^\delta(F[j])$.
  Therefore, $\W_\sys^\delta\preceq W$.
\end{IEEEproof}
\vspace{1mm}
\begin{exmp}
   Consider again Example \ref{exp:running}. We can choose $\varepsilon=\mu=0.005$ by Theorem \ref{thm:maindba} since the Lipschitz constant $\rho=1$ (using Remark \ref{rem:rho}). Let $Y=[W[2], W[1]]^T$ and $Z=W[0]$ with $W[i]$ approximating $W_\sys(q_i)$ ($i=0,1,2$) in Algorithm~\ref{alg:buchigame}. Initially, $Y^0=[\emptyset,\emptyset]^T$ and $Z^0=\X$. For the $l$th inner iteration, 
   {\footnotesize
   \begin{align*}
       W^l[1]&=W^{l-1}[1]\cup[\underline{\pre}_\mu]^\varepsilon(W^{l-1}[0]|L(a_2))\cup[\underline{\pre}_\mu]^\varepsilon(W^{l-1}[1]|L(\neg a_2)),\\
       W^l[2]&=W^{l-1}[2]\cup[\underline{\pre}_\mu]^\varepsilon(W^{l-1}[1]|L(a_1))\cup[\underline{\pre}_\mu]^\varepsilon(W^{l-1}[2]|L(\neg a_1)).
   \end{align*}}
   Algorithm~\ref{alg:buchigame} terminates after 7 inner iterations and 1 outer iteration, and it returns $W[0]=\X$, $W[1]=[0, 0.6]\cup[1.448, 2]$ and $W[2] = [0, 0.003]\cup[0.1, 0.2]\cup[1.893, 2]$, which shows (\ref{eq:maindba}).
\end{exmp}

\begin{rem}
Similar to the classic B\"uchi game algorithm designed for finite-state systems, Algorithm~\ref{alg:buchigame} is composed of two nested fixed-point iterations of $\T^\delta(\M_\varphi,\cdot)$. The outer loop computes $\win_\sys(F)$, and the inner loop computes $\win_\sys(\bar{F})$, where $\win_\sys(F)$ and $\win_\sys(\bar{F})$ denote the vectors of $\sys$-domains of states in $F$ and $\bar{F}$ ($\bar{F}=Q\setminus F$), respectively. But simply applying the classic algorithm directly to continuous-state systems (using the exact operator $\T$) will not give the actual winning set. This is because the sequence of sets $\set{Y_\nu}$ are required to be compact to conclude $\lim_{\nu\to\infty}Z_\nu=\win_\sys(F)$, which is similar to computing maximal controlled invariant sets \cite{Li2016}, while $\set{Y_\nu}$ need to be open so that $\lim_{\nu\to\infty}Y_\nu=\win_\sys(\bar{F})$, which is equivalent to the backward reachable set computation in \cite[Section 4.2]{Li2019thesis}. By using an approximation $[\T_\mu]^\varepsilon$ of $\T$, however, Algorithm~\ref{alg:buchigame} can terminate in finite time and yield an approximation of the real winning set that is lower-bounded by the winning set with $\delta$-perturbation. An arbitrarily accurate approximation of the winning set $\win_\sys(\varphi)$ can be achieved if $\lim_{\delta\to 0}\win_\sys^\delta(\varphi)=\win_\sys(\varphi)$.
\end{rem}

\begin{rem}
  With minor modification, Algorithm 1 can be changed to provide the following completeness guarantees for robust synthesis: for all $\delta_2>\delta_1\ge 0$, if $\rho(\varepsilon+\mu)\leq\delta_2-\delta_1$, then
  $$
    \win_\sys^{\delta_2}(\varphi)\subseteq W(q_0)\subseteq\win_\sys^{\delta_1}(\varphi).
 $$
\end{rem}

\subsection{Automata-embedded control structure}
Algorithm~\ref{alg:buchigame} essentially induces a finite-memory control strategy that is embedded with the given DBA.
\begin{defn}\label{def:globalctlr}
  Let $\A_\varphi=(Q,\Sigma,r,q_0,F)$ be an equivalent DBA of an LTL formula $\varphi$. An \emph{automaton-embedded control strategy} for system $\sys=\langle\X,\act,\D,R,AP,L\rangle$ is
  \begin{align}\label{eq:c}
    \ctlr_\varphi=\langle \X_c,\act_c,Q_c,\Sigma_c,r_c,q_0,H\rangle:
  \end{align}
  \begin{itemize}
  \item $\X_c\subseteq\X$ is a set of inputs;
  \item $Q_c=Q$ is a finite set of states;
  \item $\Sigma_c=\Sigma=2^{AP}$ is an alphabet;
  \item $r_c=r\subseteq Q_c\times \Sigma_c\times Q_c$ is a transition relation that updates the controller state;
  \item $q_0$ is the initial state;
  \item $\act_c\subseteq 2^\act$ is a set of outputs;
  \item $H:Q_c\times\X_c\to\act_c$ is an output function defined by
    \begin{align*}
      H(q,x)=\kappa_{Id(q)+1}(x),\; x\in\X_c, q\in Q_c,
    \end{align*}
    where $\kappa_{Id(q)}(x)$ belongs to the set of memoryless control strategies $\set{\kappa_i}_{i=1}^{\card{Q_c}}$ returned by Algorithm~\ref{alg:buchigame} and $Id(q)$ is the index of the state $q$.
  \end{itemize}
\end{defn}

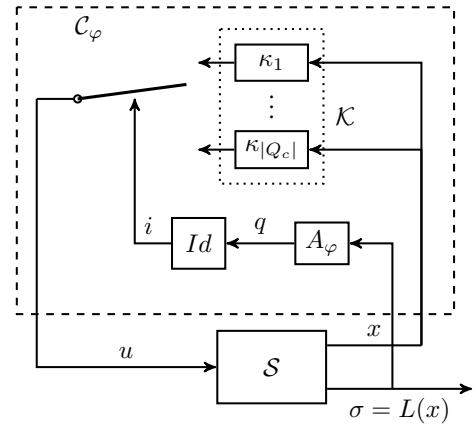
\begin{figure}[htbp]
  \centering
  \resizebox{0.7\linewidth}{!}{%
  \begin{tikzpicture}[align=center,->,>=stealth',thick]
    \node (plant) [draw,thick,minimum width=1.5cm,minimum height=1cm] {$\sys$};
    \node (kl) [above of=plant, xshift=0cm, yshift=2cm, draw,thick,minimum width=1cm,minimum height=0.5cm] {$\kappa_{\card{Q_c}}$};
    \node (kdots) [above of=kl,node distance=0.7cm] {$\vdots$};
    \node (k1) [above of=kdots,node distance=0.5cm,draw,thick,minimum width=1cm,minimum height=0.5cm] {$\kappa_1$};
    \node (output) [above left of=plant,xshift=-.3cm,yshift=1cm,draw,thick,minimum width=0.7cm,minimum height=0.7cm, text width=0.5cm] {$Id$};
    \node (automaton) [right of=output,xshift=0.7cm,draw,thick,minimum width=0.7cm,minimum height=0.5cm, text width=0.5cm] {$A_\varphi$};
    \draw [-] ($(plant.east)+(0,0.3)$) -- node [above] {$x$} ($(plant.east)+(1.3,0.3)$);
    \draw ($(plant.east)+(0,-0.3)$) -- node [below] {$\sigma=L(x)$} ($(plant.east)+(2,-0.3)$);
    \draw (k1.west) -- ($(k1.west)+(-0.5,0)$);
    \draw (kl.west) -- ($(kl.west)+(-0.5,0)$);
    \draw ($(plant.east)+(1.3,0.3)$) |- (k1.east);
    \draw ($(plant.east)+(1.3,0.3)$) |- (kl.east);
    \draw [-,very thick] ($(kdots.west)+(-1,0.2)$) -- ($(kdots.west)+(-2.5,0)$);
    \node (circle) at ($(kdots.west)+(-2.5,0)$) [circle,draw,inner sep=0pt,minimum size=1mm]{};
    \draw [-] (circle.west) -- ($(circle.west)+(-0.5,0)$) coordinate (c1);
    \draw (c1) |- node [above right,xshift=1cm] {$u$} (plant.west);
    
    \draw ($(plant.east)+(0.9,-0.3)$) |- (automaton.east);
    \draw (automaton.west) -- node [above] {$q$} (output.east);
    \draw (output.west) -| node [above right] {$i$} ($(output.west)+(-0.5,2)$);

    \draw [dotted, thick] ($(k1.north west)+(-0.2,0.2)$) rectangle ($(kl.south east)+(0.2,-0.2)$);
    \node at ($(k1.north east)+(0.5,-1)$) {$\mathcal{K}$};
    \draw [dashed, thick] ($(k1.north west)+(-3,0.5)$) rectangle ($(kl.south east)+(2,-2)$);
    \node at ($(k1.north west)+(-2,0.2)$) {$\mathcal{C}_\varphi$};
  \end{tikzpicture}
  }
  \caption{The finite-memory controller structure.}
  \label{fig:strategy}
\end{figure}

The components $Q_c,\Sigma_c,r_c,q_0$ originally given in $\A_\varphi$ are embedded into $\ctlr_\varphi$. One can use a single variable that takes values in a subset of $\N$ to represent $Q$. Such a variable is called a \emph{memory variable}. A memoryless control strategy $\kappa$ from $\K$ is activated by the function $H$, which outputs the index of current state $q$ of $\A_\varphi$ by the transition relation $r$ of $\A_\varphi$ according to the previous automaton state and the labels $L(x)$ of the current system state $x$.
Therefore, the embedded $\A_\varphi$ manages the control memory, and the structure in Definition \ref{def:globalctlr} is visualized in Fig.~\ref{fig:strategy}.

\begin{cor}
  Consider system $\sys^0$ and an LTL formula $\varphi$. Let $\A_\varphi$ be an equivalent DBA for an LTL specification $\varphi$. If $\varphi$ is robustly realizable for $\sys^0$, then there exists a finite-memory control strategy (\ref{eq:c}) to realize $\varphi$ for $\sys^0$.
\end{cor}

\begin{rem}
  It is worth noting that the output $W$ of Algorithm~\ref{alg:buchigame} is a vector of unions of non-uniform intervals that inner approximate the $\sys$-domains of $\A_\varphi$. This is a result of using an interval implementation $[\T_\mu]^\varepsilon$ of $\T$, in which the approximation of predecessors are obtained by adaptively partition the state space $\X$ w.r.t. both the dynamics and the target sets. Such a computation scheme can lead to an efficiency gain by avoiding discretizing $\X$ uniformly with a high precision except in constraint-critical areas.
\end{rem}

\subsection{Sound synthesis for full LTL}\label{sec:full+soundness}
As shown in previous sections, DBA not only guide control synthesis but also serve as the memory update mechanism in the resulting finite-memory controller. Unlike DBA, NBA cannot be used to update controller memory variables because the updating rule is non-deterministic. For this reason, determinization of NBA (into deterministic automata such as deterministic Rabin automata (DRA)) is usually applied in general LTL control synthesis. 
Similar to Algorithm 1, an interval implementation of a Rabin game algorithm \cite{piterman2006faster} would, in principle, give robust completeness guarantees for full LTL specifications. Nonetheless, the inherent complexity of a complete algorithm for solving a Rabin game would render such an approach impractical. 

In most of the control applications, it is usually unnecessary to seek a complete set of control solutions. Sound solutions that can be found in reasonable time are more appealing. A simple and direct treatment is to trim off redundant transitions under the same input propositions in an NBA, which generates a DBA that specifies a subset of behaviors of the original NBA.
Based on the DBA trimmed from the NBA translated from a general LTL formula, Algorithm~\ref{alg:buchigame} can provide sound solutions to the corresponding control synthesis problem. The following example demonstrates such a process.

\begin{exmp}
   Consider a discrete system $\sys'$ (Fig.~\ref{fig:nts}) and the correspond NBA (Fig.~\ref{fig:FGb}) of the formula $\eventually\always b$. The NBA can be trimmed into the DBA in Fig.~\ref{fig:FGb-dba}.
   $L(s_1)=L(s_2)=\set{b}$ and $L(s_{0})=L(s_{3})=L(s_{4})=\set{\neg b}$.
   \begin{figure}[htbp]
    \centering
    \begin{subfigure}[htbp]{0.5\columnwidth}
      \resizebox{0.9\linewidth}{!}{%
        \begin{tikzpicture}[->,>=stealth',auto,node distance=1.5cm,semithick]
          \node[state] (s_3) {$s_3$};
          \node[state] (s_0) [above left of=s_3] {$s_0$};
          \node[state] (s_4) [below left of=s_3] {$s_4$};
          \node[state] (s_1) [above right of=s_3] {$s_1$};
          \node[state] (s_2) [below right of=s_3] {$s_2$};
          \path (s_0) edge node {$u_1$} (s_3)
          edge [loop left] node {$u_2$} (s_0)
          edge node [above left] {$u_1$} (s_4)
          (s_4) edge node {$u_4$} (s_3)
          edge node {$u_4$} (s_2)
          (s_2) edge node [above right] {$v_2$} (s_3)
          edge [loop right] node {$v_2$} (s_2)
          (s_3) edge node {$u_3$} (s_1)
          (s_1) edge [loop right] node {$v_1$} (s_1);
        \end{tikzpicture}
      }
      \caption{$\sys'$}
      \label{fig:nts}
    \end{subfigure}%
    \begin{subfigure}[htbp]{0.5\columnwidth}
        \begin{subfigure}[htbp]{1.0\columnwidth}
        \resizebox{0.8\linewidth}{!}{%
        \begin{tikzpicture}[->,>=stealth',auto,node distance=1.5cm,semithick]
            \node[state,initial] (v_0) {$q_0$};
            \node[state,accepting] (v_1) [right of=v_0] {$q_1$};
            \node[state] (v_2) [right of=v_1] {$q_2$};
            \path (v_0) edge node {$b$} (v_1)
            edge [loop above] node {$\top$} (v_0)
            (v_1) edge [loop above] node {$b$} (v_1)
            edge node {$\neg b$} (v_2)
            (v_2) edge [loop above] node {$\top$} (v_2);
        \end{tikzpicture}
        }
        \caption{$\A'$}
        \label{fig:FGb}
        \end{subfigure}
        \begin{subfigure}[htbp]{1.0\columnwidth}
        \resizebox{0.8\linewidth}{!}{%
        \begin{tikzpicture}[->,>=stealth',auto,node distance=1.5cm,semithick]
            \node[state,initial] (v_0) {$q_0$};
            \node[state,accepting] (v_1) [right of=v_0] {$q_1$};
            \node[state] (v_2) [right of=v_1] {$q_2$};
            \path (v_0) edge node {$b$} (v_1)
            edge [loop above] node {$\neg b$} (v_0)
            (v_1) edge [loop above] node {$b$} (v_1)
            edge node {$\neg b$} (v_2)
            (v_2) edge [loop above] node {$\top$} (v_2);
        \end{tikzpicture}
        }
        \caption{The DBA trimmed from $\A'$}
        \label{fig:FGb-dba}
        \end{subfigure}
    \end{subfigure}
    \caption{The system $\sys'$, the NBA $\A'$ of $\eventually\always b$, and DBA trimmed from $\A'$.}
    \label{fig:reachstay}
    \end{figure}
The real winning set of system $\sys'$ w.r.t. $\eventually\always b$ is $\set{s_0,s_1,s_2,s_3, s_4}$. By using Algorithm~\ref{alg:buchigame}, the vector $Y$, which will be returned as a vector $[W_{\sys}(q_0), W_{\sys}(q_2)]^T$ of $\sys$-domains, is initialized to $Y_0^0=[\emptyset,\emptyset]^T$, and $Z$, which represents $W_{\sys}(q_1)$, is initialized to $Z_0=\set{s_0,s_1,s_2,s_3, s_4}$. By the transition matrix of $\A'$ and (\ref{eq:T}), for the $l$th iteration of the $1$st inner loop (line 8-12), $Y_0^l[1]=Y_0^{l-1}\cup\pre^\delta(Z_0|L^{-1}(b))\cup \pre^\delta(Y_0^{l-1}[1]|L^{-1}(\neg b))=\set{s_1, s_2}$, and the $1$st inner loop finishes in 4 iterations with $Y_0[1]=Y_0^4[1]=\set{s_0,s_1,s_2,s_3, s_4}$ and $Y_0[2]=\emptyset$. Then $Z_1=\pre^\delta(Z_0|L^{-1}(b))\cup \pre^\delta(Y_0[2]|L^{-1}(\neg b))=\set{s_1,s_2}$. Using $Z_1$ in the $2$nd inner loop gives $Y_1[1]=\set{s_1,s_3}$ and $Y_1[2]=\emptyset$. As such, the returned vectors are $Z=\set{s_1}$ and $Y=[\set{s_1, s_3}, \emptyset]^T$. Since $q_0$ is the initial node, the winning set obtained by Algrithm~\ref{alg:buchigame} is $Y[1]=\set{s_1, s_3}$, which is a subset of the real one.
\end{exmp}

We also demonstrate in Section \ref{sec:evaluation} with an application of jet engine compressor control that such a sound solution is sufficient to have a satisfactory control synthesis result.

\section{Control Synthesis with Pre-processing}\label{sec:preprocess}

We are also concerned with computational complexity of control synthesis. In Algorithm~\ref{alg:buchigame}, the vector $Z$ updates only after the vector $Y$ remains unchanged in the inner loop. In addition, at the beginning of each computation in the outer loop, the value of $Y$ needs to be reinitialized since the value of $Z$ is changed from the last iteration. In this sense, the interdependency between $Z$ and $Y$ increases the complexity.

Suppose that the transition matrix $\M_\varphi$ of a DBA $\A_\varphi$ is in the form:
\vspace{-2mm}
\begin{align} \label{eq:Mblocks}
  \M_\varphi=\begin{bmatrix}\M_{11} & \M_{12}\\ \M_{e} & \M_{22}\end{bmatrix}_{\card{Q}\times\card{Q}},
\end{align}
where $\M_{11}$ and $\M_{22}$ are $n_L$ by $n_L$ and $n_R$ by $n_R$ matrices, respectively, and $n_L+n_R=\card{Q}$, $n_L,n_R\in\Zp$. $\M_{e}$ is a matrix of empty symbols (i.e., $e$'s). Let $Q_L$ be the set of states of $\A_\varphi$ with the first $n_L$ indices and $Q_R$ be the set of the rest of the states. Define $F_L=\set{q\in Q\sv q\in F \wedge q\in Q_L}$ and $F_R=\set{q\in Q\sv q\in F \wedge q\in Q_R}$.
Denote $n_{L2}=\card{F_L}$, $n_{R2}=\card{F_R}$, $n_{L1}=n_L-n_{L2}$, and $n_{R1}=n_R-n_{R2}$. We also assume that the states in $Q_L$ and $Q_R$ are sorted so that the accepting states always rank after nonaccepting ones.

If $Q_R$ contains accepting nodes, then $\M_{22}$ can be treated as a sub-transition matrix based on which $\set{\W_\sys(q)}_{q\in Q_R}$ can be approximated firstly by Algorithm~\ref{alg:buchigame}, independent of other parts of $\A_\varphi$. If $Q_R$ has no accepting nodes, then computing $\set{\W_\sys(q)}_{q\in Q_R}$ is pointless because there is no transition from any $q\in Q_R$ to $q'\in Q_L$ and any run that contains $q$ does not satisfy the B\"uchi accepting condition. The approximation of $\set{\W_\sys(q)}_{q\in Q_L}$, starts after the computation w.r.t. $\M_{22}$ completes. In this way, the repetitive initialization and computation of $W_L$ caused by the updates in $W_R$ can be avoided. Therefore, if we can arrange the transition matrix $\M_\varphi$ into a triangular matrix or triangular block matrix without changing the original transition relations in $\A_\varphi$, then Algorithm~\ref{alg:buchigame} can reduce to a single loop or several smaller nested loops. We now compare the complexities of Algorithm~\ref{alg:buchigame} applying directly to $\M_\varphi$ and sequentially to the blocks in (\ref{eq:Mblocks}).

Suppose that the number of accepting and nonaccepting nodes in $\A_\varphi$ is $n_2$ and $n_1=\card{Q}-n_2$, respectively, and the resulting numbers of outer-loop and inner-loop iterations by using Algorithm~\ref{alg:buchigame} directly are $K_2$ and $K_1$. Then the complexity is $\mathcal{O}(n_2K_2n_1K_1)$ for the control synthesis without using its triangular form. Let the numbers of outer and inner-loop iterations for block $\M_{22}$ be $K_{R2}$ and $K_{R1}$, respectively, and the ones for block $\begin{bmatrix}\M_{11}&\M_{12}\end{bmatrix}$ be $K_{L2}$ and $K_{L1}$, respectively. If we perform control synthesis sequentially to blocks $\M_{22}$ and $\begin{bmatrix}\M_{11}&\M_{12}\end{bmatrix}$, the complexity is $\mathcal{O}(n_{R2}K_{R2}n_{R1}K_{R1}+n_{L2}K_{L2}n_{L1}K_{L1})$.
The number of outer and inner-loop iterations are determined by the row that converges the slowest.
As defined before, $n_2=n_{L2}+n_{R2}$ and $n_1=n_{L1}+n_{R1}$. Then
\begin{align*}
  &n_{R2}K_{R2}n_{R1}K_{R1}+n_{L2}K_{L2}n_{L1}K_{L1}\\
  &\leq (n_{R2}n_{R1}+n_{L2}n_{L1})K_2K_1\\
  &<(n_{R2}+n_{L2})(n_{R1}+n_{L1})K_2K_1=n_2n_1K_2K_1,
\end{align*}
which shows that we can gain computational efficiency by using an upper triangular block matrix.

To reform $\M_\varphi$ so that $\M_\varphi$ is a triangular block matrix, we propose the following \textsc{Preprocess} procedure.

\begin{algorithm}
  \caption{$\M_\varphi=\textsc{Preprocess}(\A_\varphi)$}
  \label{alg:sdomextra}
  \begin{algorithmic}[1]
    \State Detect all SCCs\footnotemark in the graph representation of $\A_\varphi$. Then $\A_\varphi$ is simplified to a DAG\footnotemark $\graph_{dag}=(V,E)$ in which each node $v\in V$ is either a single state or an SCC.
    \State Perform a \emph{topological sort} on the DAG $\graph_{dag}$, which determines a linear ordering of the nodes in $\graph_{dag}$ so that $v$ precedes $v'$ for any $(v,\sigma,v')\in E$. Let $v_1\cdots v_k \cdots v_{\card{V}}$ ($1\leq k\leq \card{V}$) be the resulting order and $v_k$ is the last node that is or contains an accepting state.
    \State List the states in $Q$ in the order of $v_1\cdots v_k$. No specific order of the states in the same $v\in V$ is required but the accepting states rank after the nonaccepting ones.
    \State Write $\M_\varphi$ w.r.t. the current order.
  \end{algorithmic}
\end{algorithm}
\footnotetext[1]{A strongly connected component (SCC) is a (sub)graph where there exists a path between any two nodes. An SCC with one node is called a trivial SCC.}
\footnotetext[2]{A directed acyclic graph (DAG) is a directed graph without cycles.}

The transition matrix $\M_\varphi$ based on the order of the automaton states obtained by \textsc{Preprocess} can be formulated as an upper triangular block matrix. Algorithm~\ref{alg:buchigame}, as a result, can be performed independently for the sub-matrices in the reversed order of the topological sort.

\begin{exmp}\label{eg:demo}
  Consider the LTL formula $\varphi_1=\eventually(a_1\wedge\eventually(a_2\wedge\eventually(a_3\wedge(\neg a_2)\until a_1)))$. Its translated DBA is shown in Fig.~\ref{fig:robotltl}.
  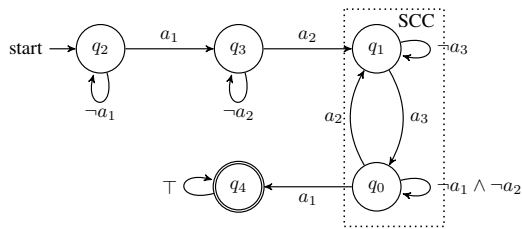
\begin{figure}[htbp]
    \centering
    \resizebox{0.8\linewidth}{!}{%
    \begin{tikzpicture}[->,>=stealth',auto,node distance=2.5cm,semithick]
      \node[state,initial] (v_0) {$q_2$};
      \node[state] (v_3) [right of=v_0] {$q_3$};
      \node[state] (v_2) [right of=v_3] {$q_1$};
      \node[state] (v_4) [below of=v_2] {$q_0$};
      \node[state,accepting] (v_1) [left of=v_4] {$q_4$};
      \path (v_0) edge [loop below] node {$\neg a_1$} (v_0)
      edge node {$a_1$} (v_3)
      (v_3) edge [loop below] node {$\neg a_2$} (v_3)
      edge node {$a_2$} (v_2)
      (v_2) edge [loop right] node {$\neg a_3$} (v_2)
      edge [bend left] node {$a_3$} (v_4)
      (v_4) edge [loop right] node {$\neg a_1\wedge\neg a_2$} (v_4)
      edge [bend left] node {$a_2$} (v_2)
      edge node {$a_1$} (v_1)
      (v_1) edge [loop left] node {$\top$} (v_1);
      \draw [dotted,thick] ($(v_2.north west)+(-0.27,0.4)$) rectangle ($(v_4.south east)+(0.9,-0.4)$);
      \node at ($(v_2.north east)+(0.4,0.2)$) {SCC};
    \end{tikzpicture}
    }
    \caption{The translated DBA using Spot \cite{Duret-Lutz2016}.}
    \label{fig:robotltl}
  \end{figure}
  
  The states $q_1$ and $q_0$ constitute an SCC $q_{1,0}$, and the rest of the states are trivial SCCs. A topological sort of $\A_\varphi$ is $q_2q_3q_{1,0}q_4$, and $q_4$ is the unique accepting state. Then the \textsc{Preprocess} yields an order of the states in $Q$: $q_2q_3q_1q_0q_4$.
  Based on this order, the transition matrix is
  \begin{align*}
    \M_{\varphi_1}=\left[
    \begin{array}{cc:cc:c}
      \neg a_1&\multicolumn{1}{:c:}{a_1}&e&e&e\\\cdashline{1-4}
      e&\multicolumn{1}{:c:}{\neg a_2}&a_2&e&e\\\hdashline
      e&e&\neg a_3&a_3&e\\
      e&e&a_2&\neg a_1\wedge\neg a_2& a_1\\ \cdashline{3-5}
      e&e&e&e&\top
    \end{array}\right]
  \end{align*}

  Since $q_4$ is the only accepting state and its corresponding block matrix $\M_4=\top$, $\W_\sys(q_4)=\X$ and $\kappa_4(x)=\act$ for all $x\in\X$. Algorithm~\ref{alg:buchigame} is then applied to $\M_\varphi$ backwardly and blockwisely.
\end{exmp}

\section{Performance Evaluation}\label{sec:evaluation}

\subsection{Complexity analysis}
By using a branch-and-bound scheme (Algorithm \ref{alg:cpre}) in Algorithm \ref{alg:buchigame}, the discretization of the state space $\X$ is only refined in local areas regarding the satisfaction of the DBA. As a result, the size of the discretized system can be reduced. In addition, no space is required to store the abstraction and the product system with a possibly huge number of transitions.
These all lead to a lower space complexity for control synthesis. On the other hand, managing nonlinear data structure in the proposed method induces overhead cost in time, which makes the time complexity of the proposed method higher than abstraction-based methods in the worst case.
In this section, we provide the complexity analysis of Algorithm \ref{alg:buchigame} and compare it with the abstraction-based methods.

Assume that $\varepsilon>0$ is the uniform grid size of the state space $\X$ for abstraction-based methods and the minimum width of an interval in Algorithm~\ref{alg:cpre}. For both methods, $\mu>0$ is the under-sample parameter of the control space $\act$ of system $\sys$.
Then the number of states and controls in the abstraction are of $N_\X=\mathcal{O}((1/\varepsilon)^n)$ and $N_\act=\mathcal{O}((1/\mu)^m)$, respectively. In the abstraction, there is a transition from an interval (or a grid cell) $[x]$ to an interval $[x']$ as long as there exists a control $u\in[\act]_\mu$ such that the one-step reachable set of $[x]$ under $u$ intersects with $[x']$. By Assumption \ref{asp:lipschitz}, such a reachable set can be $\rho^n$ times the volume of a single interval for system $\sys$, and therefore the reachable set can intersect with as much as $(\ceil{\rho}+1)^n$ intervals. By enumerating all intervals and control inputs, we have $N_R=\mathcal{O}((\ceil{\rho}+1)^n N_\X N_\act)$ as the number of transitions. The product of an abstraction of system $\sys$ and the DBA $\A_\varphi$ has $N_\X^{\rm prod}=\mathcal{O}(\abs{Q}N_\X)$ number of states and $N_R^{prod}=\mathcal{O}(\abs{Q}N_R)$ number of transitions. 
For the proposed method, Algorithm~\ref{alg:cpre} can be implemented by using a binary tree, and the operator $[\T_\mu]^\varepsilon$ in Algorithm~\ref{alg:buchigame} requires $\abs{Q}$ such binary trees as defined in (\ref{eq:T}). The branch-and-bound scheme usually results in an unbalanced tree, but in the worst case, each binary tree is balanced and has $N_\X$ number of leaves, which means that the tree height is $h=\log_2(N_\X)$ and there are at most $\sum_{i=0}^h 2^i=2N_\X-1$ nodes.

For abstraction-based methods, the space complexity is proportional to the number of transitions in the product of the abstraction and the DBA, and the computational time includes the time for abstraction and control synthesis, which are $\mathcal{O}(N_\X^{\rm prod})$ and $\mathcal{O}(N_\X^{\rm prod}N_R^{prod})$, respectively. For the proposed method, the space complexity is $\mathcal{O}(2\abs{Q}N_\X)$ since $\abs{Q}$ binary trees are constructed. To analyze the time complexity of Algorithm \ref{alg:buchigame}, we assume the worst case that $Y$ and $\widetilde{Y}$ only differ in one of the elements (i.e., $Y[i]$ for some $i\in \set{1,\cdots,n_1}$) by one interval in the inner loop and the same for the outer loop. Hence, the maximal numbers of inner and outer iterations are $(\abs{Q}-\abs{F})N_\X$ and $\abs{F}N_\X$, respectively. Note that the complexity of a single loop in Algorithm \ref{alg:cpre} is $\mathcal{O}((\log_2 N_\X+c)N_\act)$, which includes the membership test of an interval along the tree and the cost of computing $[y]=[f]([x],u)$ relative to the membership test (denoted by a factor $c$).
Then the complexity of Algorithm~\ref{alg:cpre}, assuming there are $N_\X$ number of iterations, is $N_{\pre}=\mathcal{O}(N_\X(\log_2 N_\X+c)N_\act)$. The complexity for the operator $[\T_\mu]^\varepsilon$ is then $\mathcal{O}(\abs{Q}N_{\pre})$ by definition. Therefore, the overall time complexity of Algorithm \ref{alg:buchigame} is $\mathcal{O}(((\abs{Q}-\abs{F})N_\X+1)\abs{F}N_\X \abs{Q}N_{\pre})=\mathcal{O}(\abs{Q}^2\abs{F}N_\X^3(\log_2N_\X+c)N_\act)$, considering $(\abs{Q}-\abs{F})N_\X+1\approx\abs{Q}N_\X$.

We summarize the above analysis in Table \ref{tab:complexity}, which shows that the proposed method is more efficient in memory usage: usually $(\ceil{\rho}+1)^n N_\act\gg 2$. The main reason is that it does not store all the transitions of the product system, which can be exceptionally huge when $\varepsilon$ needs to be very small or the dimension of system $\sys$ is high. The abstraction-based methods has a lower time complexity than the worst case of the proposed method. This is because the proposed method trades time for saving the space to store transitions.
However, the worst case rarely happen in practice.
\begin{table}[ht]
\centering
  \caption{Complexity comparison: abstraction-based methods (first row) v.s. Algorithm~\ref{alg:buchigame} (second row).}
  \begin{tabular}{cc}
    \toprule
    Time complexity & Space complexity \\
    \hline
    $\mathcal{O}(\abs{Q}N_\X+(\ceil{\rho}+1)^n \abs{Q}^2N_\X^2N_\act)$ & $\mathcal{O}((\ceil{\rho}+1)^n N_\act\abs{Q}N_\X)$\\ 
    $\mathcal{O}(\abs{Q}^2\abs{F}N_\X^3(\log_2N_\X+c)N_\act)$ & $\mathcal{O}(2\abs{Q}N_\X)$\\
    \bottomrule
  \end{tabular}
  \label{tab:complexity}
\end{table}

Since both abstraction-based methods and the proposed method operates on a discretized state space, the computational complexity is exponential in terms of the system dimension, which is an inherent shortcoming of such discretization-based control methods.

\begin{rem}
The complexity of applying the controller synthesized by the proposed method on the fly is $\mathcal{O}(\log_2N_\X)$ because the control value for the current system state is searched along the binary tree. One way to reduce this complexity is to map the binary tree to a uniform partition of the state space (with the granularity as the size of the smallest interval of the binary tree) through which online querying of valid control values is a constant time.
\end{rem}

\subsection{Case study}
We now demonstrate the effectiveness of Algorithm~\ref{alg:buchigame} on three different control systems and compare the performance between abstraction-based methods and the proposed specification-guided method. LTL to DBA translations in all cases are obtained by using Spot \cite{Duret-Lutz2016}, and the edges of the DBAs are simplified according to the actual labeling function. The experimental results presented here are obtained by Algorithm~\ref{alg:buchigame} implemented based on ROCS \cite{Li2018rocs}, and the source code can be found in https://git.uwaterloo.ca/hybrid-systems-lab/rocs.

\subsubsection{Moore-Greitzer engine control}
In this first case, we consider a reach-avoid-stay control problem for the Moore-Greitzer ODE model of a jet engine. Such a control problem is driven by the need of switching between engine operation points while avoiding entering regions of low average pressure. The following Moore-Greitzer model is used:
\begin{align}\label{eq:MGengine}
  \begin{cases}
    \dot{\Phi}=\frac{1}{l_c}(\psi_c-\Psi)+u,\\
    \dot{\Psi}=\frac{1}{4l_cB^2}(\Phi-\mu\sqrt{\Psi}),
  \end{cases}
\end{align}
where $\psi_c=a+H[1+1.5(\Phi/W-1)-0.5(\Phi/W-1)^3]$, $a=1/3.5$, $H=0.18$, $l_c=8$, $B=2$, $W=0.25$ are engine parameters related to the configuration. The states $\Phi$ and $\Psi$ are the average flow rate and pressure of an axial-flow jet engine compressor, respectively. The control inputs are the throttle coefficient $\mu$ and an additional control $u$.

\begin{figure}[htbp]
    \centering
    \begin{subfigure}[htbp]{0.5\linewidth}
      \includegraphics[width=1\linewidth]{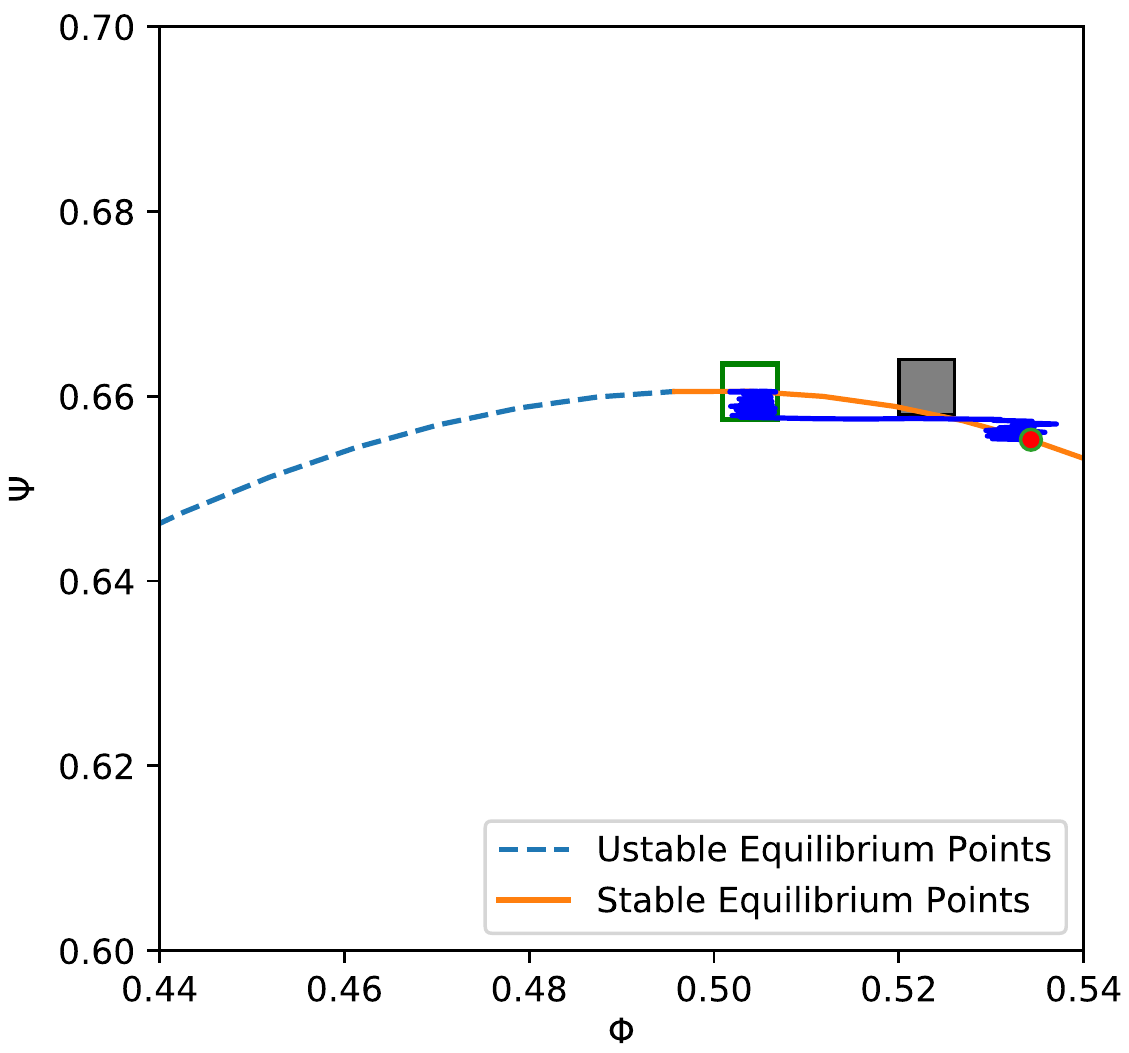}
      \caption{Scenario $\always\neg o_1\wedge\eventually\always b_1$}
    \end{subfigure}%
    \begin{subfigure}[htbp]{0.5\linewidth}
      \includegraphics[width=1\linewidth]{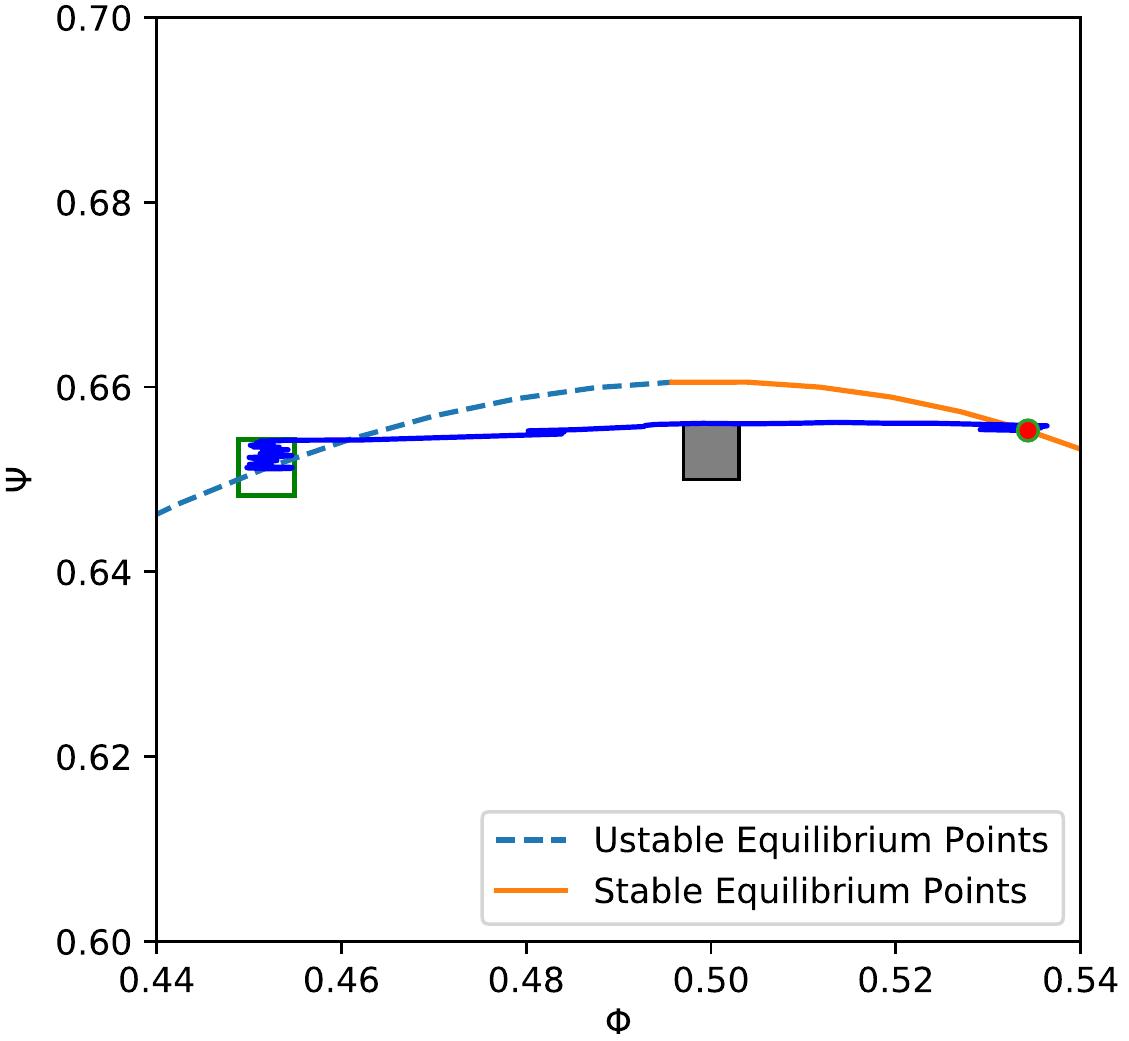}
      \caption{Scenario $\always\neg o_2\wedge\eventually\always b_2$}
    \end{subfigure}
    \caption{Controlled phase portraits for both scenarios: the green boxes are targets $B_{1,2}$, and gray boxes are avoided areas $A_{1,2}$.}
    \label{fig:MG}
\end{figure}

The reach-avoid-stay specification can be written as the LTL formula $\varphi_{\rm rs}=\always \neg o \wedge \eventually\always b$, which can only be translated into an NBA (see Fig.~\ref{fig:FGb}). 
We consider two scenarios: $L^{-1}(b)=B_1=[0.5009, 0.5069]\times[0.6575,6635]$, $L^{-1}(o)=A_1=[0.520, 0.526]\times[0.658, 0.664]$, and $L^{-1}(b)=B_2=[0.4489,0.4549]\times[0.6483, 0.6543]$, $L^{-1}(o)=A_2=[0.497, 0.503]\times[0.650, 0.656]$. The target areas $B_1$ and $B_2$ are two neighborhoods around a stable equilibrium point $(0.5039,0.6605)$ and an unstable equilibrium point $(0.4513, 0.6513)$, respectively.
In both scenarios, the initial condition is $(0.5343, 0.6553)$, the state space is $X=[0.44, 0.6]\times[0.54, 0.7]$, and the control space is $U=\set{(u, \mu)\mid u\in[-0.05, 0.05], \mu\in[0.5, 0.8]}$. 
A method based on the Taylor model \cite{Li2022} is used with a sampling time $\tau_s=0.1$s to approximate the predecessors.

The approximated winning sets with precision $\num{1.8e-4}$ for both scenarios obtained by DBA control synthesis cover $\geq 99.6\%$ of the state space, which is the same as the results obtained by using the complete coB\"uchi algorithm. As shown in Fig.~\ref{fig:MG}, the system state for both scenarios can be controlled from the given initial condition to the target sets.

\subsubsection{SCARA manipulator}
\begin{figure}[htbp]
    \centering
    \includegraphics[width=0.6\columnwidth]{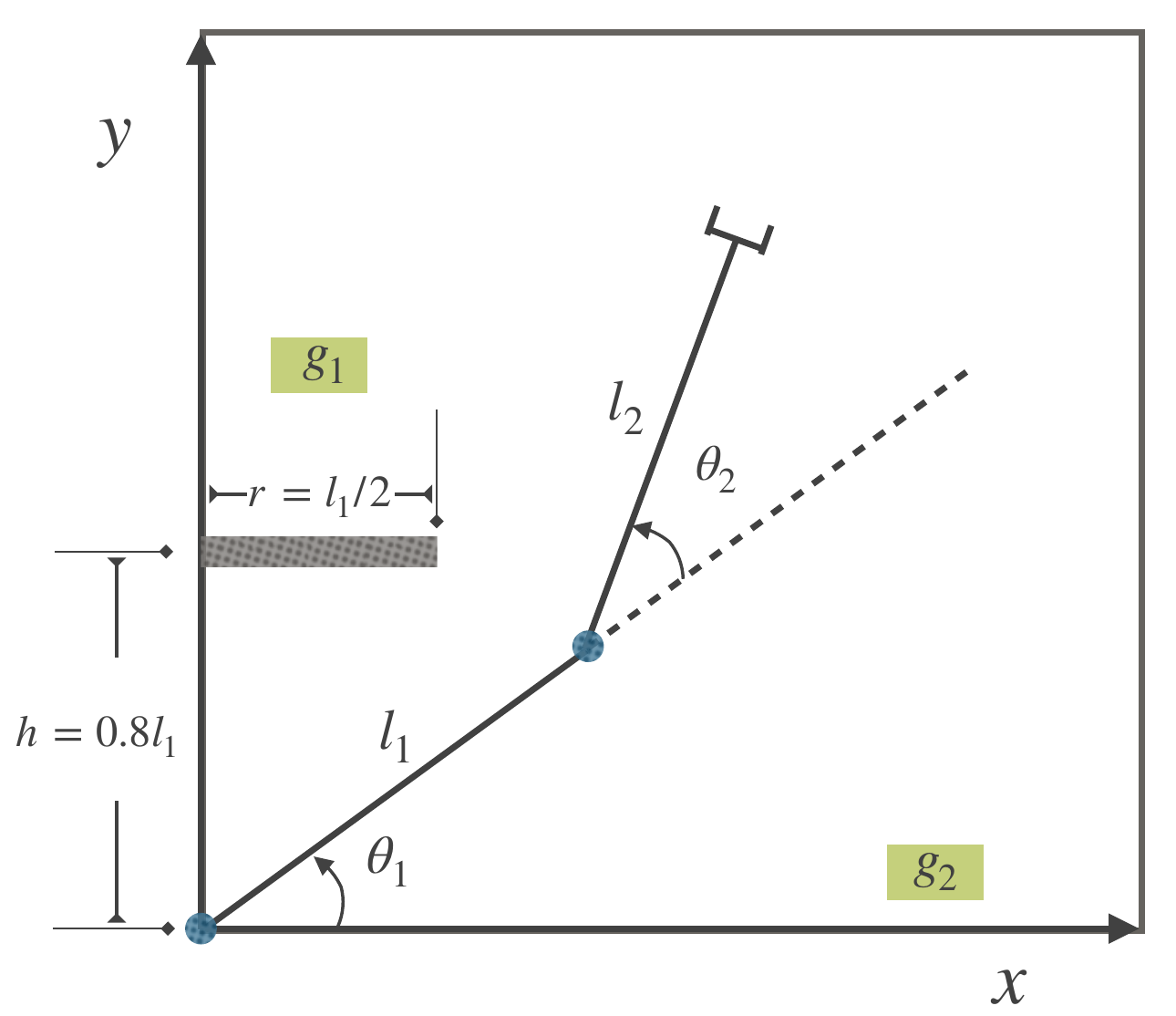}
    \caption{A two-link SCARA manipulator.}
    \label{fig:scara}
\end{figure}

SCARA (Selective Compliant Articulated Robot for Assembly) is a type of manipulators that are operate on a horizontal plane. They are often used for vertical assembly tasks in industry \cite{Siciliano2010RoboticsBook}. 
In this example, we consider a two-link SCARA manipulator in a workspace shown in Fig.~\ref{fig:scara}. Its back and fore arms are of equal length and weight ($l_1=l_2=\SI{0.15}{\metre}$, and $m_1=m_2=\SI{0.1}{\kilo\gram}$), and their moment of inertia are $I_1=I_2=\SI{1.33e-5}{\kilo\gram\metre\squared}$. There are two joints controlling the rotation of the arms, and their angles are denoted by $\theta_1$ and $\theta_2$, respectively. While moving in the workspace, the manipulator has to avoid an obstacle of length $r=0.5l_1$ at $h=0.8l_1$ to the origin.

The end-effector mounted at the end of the fore arm is required to visit area $g_1$ and $g_2$ infinitely often while avoiding the horizontal bar $o$, i.e., $\varphi_{\rm gb}=\always\neg o \wedge \always\eventually g_1\wedge\always\eventually g_2$. The classic approach to solving such a control problem is to design the trajectories that satisfy the geometry constraints and a tracking controller independently. Since the constraints are dealt with only in trajectory generation, there is no guarantee that no collision will occur during tracking. We show here a provably correct control design such that the specification can be satisfied without collision.

Consider the Lagrange dynamics of the manipulator \cite{Murray1994ManipulatorBook}:  
\begin{align}\label{eq:scara}
   \begin{bmatrix}
     z_1+2z_2c_2& z_3+z_2c_2\\
     z_3+z_2c_2 & z_3\\
   \end{bmatrix}
   \begin{bmatrix}
     \ddot{\theta}_1\\
     \ddot{\theta}_2\\
   \end{bmatrix} &+\nonumber\\
   \begin{bmatrix}
     -z_2s_2\dot{\theta}_2& -z_2s_2(\dot{\theta}_1+\dot{\theta}_2)\\
     z_2s_2\dot{\theta}_1 & 0\\
   \end{bmatrix}
   \begin{bmatrix}
     \dot{\theta}_1\\
     \dot{\theta}_2\\
   \end{bmatrix}
   &=\begin{bmatrix}\tau_1\\ \tau_2\end{bmatrix},
\end{align}
where the control inputs $\tau_{1,2}$ are the torques at the joints, $r_{1,2}=0.5l_{1,2}$ are the centers of mass of the arms to the joints, $c_2=\cos(\theta_2)$, $s_2=\sin(\theta_2)$, and
\begin{align*}
  z_1&=I_1+I_2+m_1r_1^2+m_2(l_1^2+r_2^2),\\
  z_2&=m_2l_1r_2,\\
  z_3&=I_2+m_2r_2^2.
\end{align*}
Let the state of the system be $[\theta_1,\theta_2,\omega_1,\omega_2]^T$, where $\omega_i$ is the angular velocity of $\theta_i$ ($i=1,2$). Written in the classic first-order form, (\ref{eq:scara}) becomes
\begin{align}\label{eq:scara2}
\begin{cases}
  \dot{\theta}_1=\omega_1,\\
  \dot{\theta}_2=\omega_2,\\
  \dot{\omega}_1=\frac{z_3\tau_1+z_2z_3s_2(2\omega_1+\omega_2)\omega_2+(z_3+z_2c_2)(z_2\omega_1^2s_2-\tau_2)}{z_3(z_1-z_3)-z_2^2c_2^2},\\
  \dot{\omega}_2=\frac{(z_1+2z_2c_2)(\tau_2-z_2\omega_1^2s_2)-(z_3+z_2c_2)(\tau_1+z_2s_2(2\omega_1+\omega_2)\omega_2)}{z_3(z_1-z_3)-z_2^2c_2^2}.
\end{cases}
\end{align}

Approximating the predecessors based on (\ref{eq:scara2}) will induce a large approximation error and increase the computational complexity due to the complex expressions. To synthesize a controller within tolerable time, we replace (\ref{eq:scara2}) by two double integrators:
\begin{align}\label{eq:dbint}
    \begin{cases}
      \dot{\theta}_1=\omega_1,\quad \dot{\omega}_1=u_1,\\
      \dot{\theta}_2=\omega_2,\quad \dot{\omega}_2=u_2,\\
    \end{cases}
\end{align}
where $u_{1,2}$ are virtual control inputs, and (\ref{eq:scara}) can be used to convert them to the real inputs $\tau_{1,2}$.

\begin{figure}[ht]
  \centering
  \begin{minipage}{0.5\columnwidth}
    \resizebox{1\linewidth}{!}{%
    \begin{tikzpicture}[->,>=stealth',auto,node distance=2.5cm,semithick,baseline=-30pt]
      \node[state] (v_1) {$q_1$};
      \node[state,initial,accepting] (v_0) [below left of=v_1] {$q_0$};
      \node[state] (v_2) [below right of=v_1] {$q_2$};
      \path (v_0) edge node {$g_2$} (v_2)
      edge node {$\neg (g_2\vee o)$} (v_1)
      (v_1) edge [loop right] node {$\neg (g_2\vee o)$} (v_1)
      edge node {$g_2$} (v_2)
      (v_2) edge [loop right] node {$\neg (g_1\vee o)$} (v_0)
      edge [bend left] node {$g_1$} (v_0);
    \end{tikzpicture}}
  \end{minipage}%
  \resizebox{0.5\linewidth}{!}{%
  \begin{minipage}{0.5\columnwidth}
  \begin{align*}
    \M_{\varphi_{\rm gb}}=
    \begin{bmatrix}
      \neg(g_2\vee o) & g_2 & e\\
      e & \neg(g_1\vee o) & g_1\\
      \neg(g_2\vee o) & g_2 & e\\
    \end{bmatrix}
  \end{align*}
  \end{minipage}}
  \caption{The DBA of $\varphi_{\rm gb}$ and the transition matrix with the order $q_1,q_2,q_0$.\vspace{-3mm}}
  \label{fig:gb2}
\end{figure}
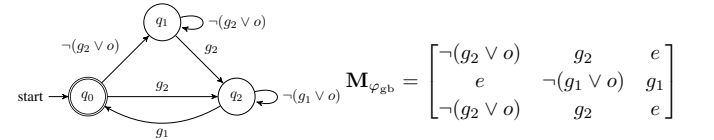

As shown in Fig.~\ref{fig:scara}, the angles for the arms satisfy $\theta_1\in[0,\pi/2]$, $\theta_2\in[-\pi,\pi]$. The collision area in the Cartesian operational space is translated into the following inequalities related to the system state:
{\small
\begin{align}
    \theta_1&\geq \arctan\left(\frac{h}{r}\right),\nonumber\\
    0\leq \theta_1\leq&\arcsin\left(\frac{h}{l_1}\right)\Rightarrow \nonumber\\ 
    \theta_1+\theta_2&\geq \pi-\arctan\left(\frac{h-l_1\sin\theta_1}{l_1\cos\theta_1-r}\right),\label{eq:collision}\\
    \arcsin\left(\frac{h}{l_1}\right)\leq\theta_1\leq &\arctan\left(\frac{h}{r}\right)\Rightarrow \nonumber\\
    \theta_1+\theta_2&\geq\pi+ \arctan\left(\frac{l_1\sin\theta_1-h}{l_1\cos\theta_1}\right). \nonumber
\end{align}
}
With \SI{0.1}{\second} as the sampling time for the discrete-time model of (\ref{eq:dbint}), the constant $\rho=1.1$ satisfies Assumption \ref{asp:lipschitz}. A discretization precision $0.05$ is used in Algorithm~\ref{alg:buchigame} for $\theta_{1,2}$ and $0.1$ for $\omega_{1,2}$. The DBA and transition matrix $\M$ of the specification $\varphi_{\rm gb}$ is given in Fig.~\ref{fig:gb2}, where propositions are simplified since the target areas are non-overlapping (same in Fig.~\ref{fig:gb}, \ref{fig:greach}, and \ref{fig:phi4}).

\begin{figure}[htbp]
    \centering
    \begin{subfigure}[htbp]{0.5\linewidth}
      \includegraphics[width=1\linewidth]{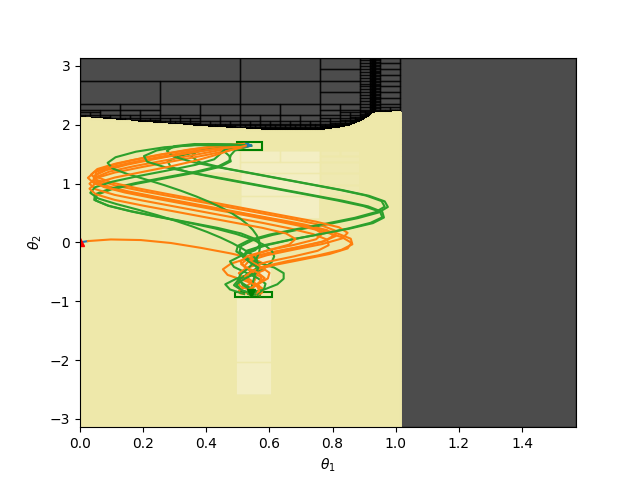}
      \caption{Trajectories in the joint space.}
    \end{subfigure}%
    \begin{subfigure}[htbp]{0.5\linewidth}
      \includegraphics[width=1\linewidth]{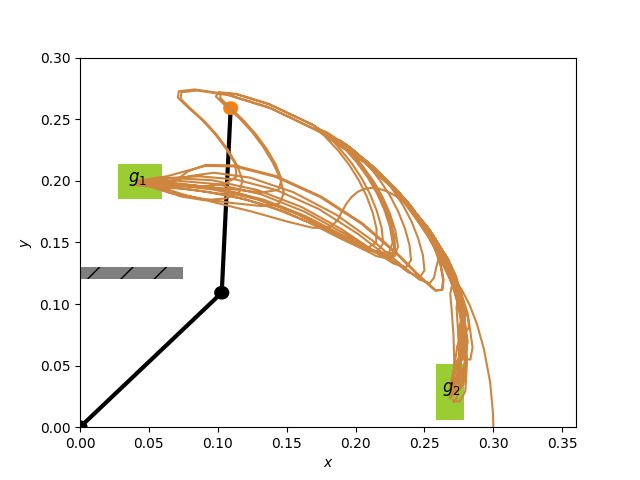}
      \caption{Trajectories in the workspace.}
    \end{subfigure}
    \begin{subfigure}[htbp]{0.5\linewidth}
      \includegraphics[width=1\linewidth]{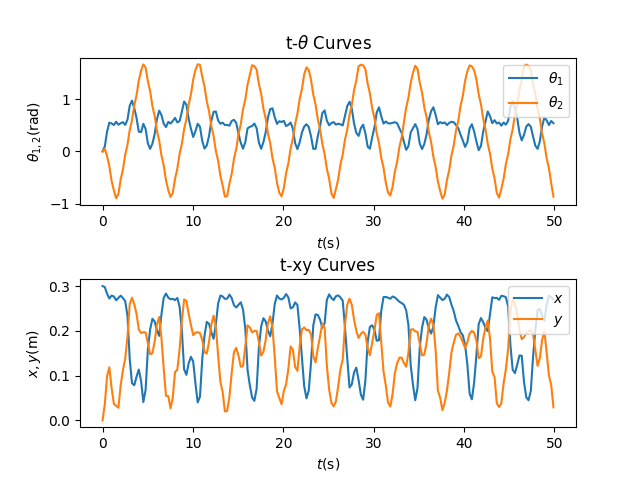}
      \caption{Time-state history.}
    \end{subfigure}%
    \begin{subfigure}[htbp]{0.5\linewidth}
      \includegraphics[width=1\linewidth]{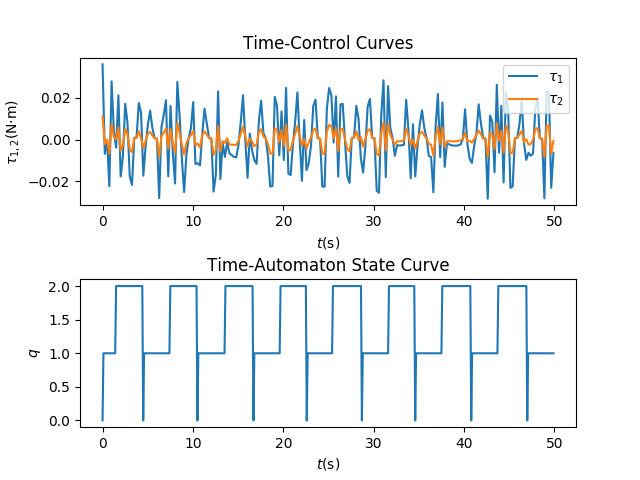}
      \caption{Time-control history.}
    \end{subfigure}
    \caption{Closed-loop simulation results. In (a), the gray area is the collision area characterized by (\ref{eq:collision}); the target sets $g_{1,2}$ are inner-approximated in the joint space by two green boxes; the winning set is projected into yellow region of the $\theta_1$-$\theta_2$ plane.}
    \label{fig:scara-trajs}
\end{figure}

As shown in Fig.~\ref{fig:scara-trajs}, the controlled trajectories by using the synthesized feedback controller are collision-free and the DBA state is switching periodically between $q_0$, $q_1$, and $q_2$. The trajectories connecting region $g_1$ and $g_2$ are not exactly the same for different periods. This is because a random valid control input is chosen when there are multiple valid control synthesized by the algorithm.

\subsubsection{Car-like mobile robot}
 We demonstrated in this section that the proposed method is well suited to solving motion planning problems. The kinematics model of the mobile robot is taken from \cite{AstromM08}, which is
\begin{equation} \label{eq:2drobot}
  \begin{bmatrix}
    \dot{x}\\\dot{y}\\\dot{\theta}
  \end{bmatrix}
  =
  \begin{bmatrix}
    v \cos (\gamma+\theta) \cos (\gamma)^{-1}\\
    v \sin (\gamma+\theta) \cos (\gamma)^{-1}\\
    v \tan (\phi)
  \end{bmatrix},
\end{equation}
where $(x,y)$ is the planar position of center of the vehicle, $\theta$ is its orientation, the control variable $v$ and $\phi$ is the velocity and steering angle, respectively, and $\gamma=\arctan (\tan (\phi)/2)$.

We use an exact discrete-time model of (\ref{eq:2drobot}) with sampling time $\tau_s=\SI{0.3}{\second}$ for the control synthesis. We consider the state space $\X=[0,10]\times[0,10]\times[-\pi,\pi]$ and the control space $\act=[-1,1]\times[1,1]$ with the sample grid $\mu=0.3$. A state space discretization precision $\leq 0.2$ in each dimension can yield winning sets that cover $\geq \SI{84}{\percent}$ of the state space.

\begin{figure}[htbp]
    \centering
    \begin{subfigure}[htbp]{0.5\linewidth}
    \includegraphics[width=1\linewidth]{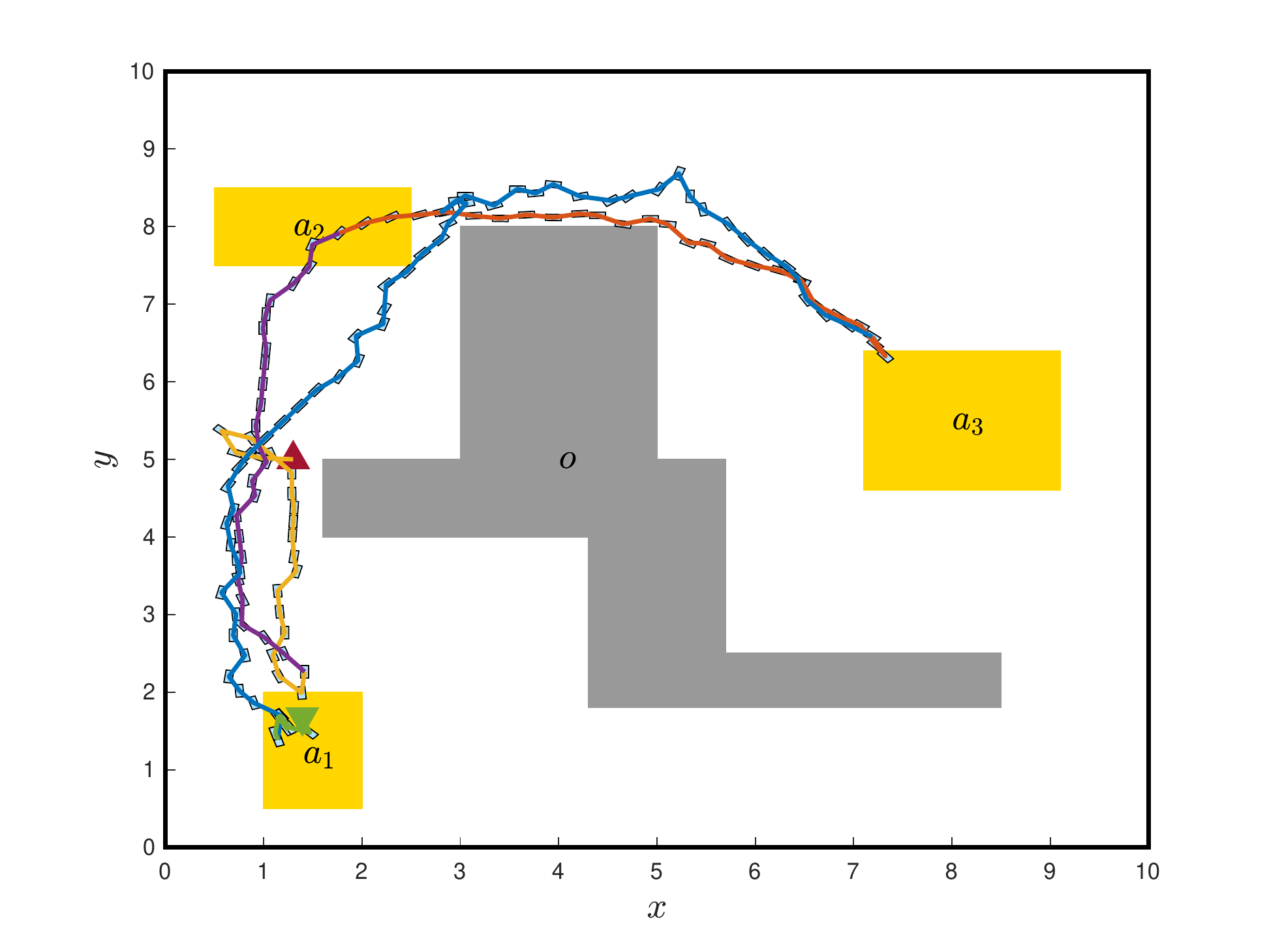}
    \caption{Closed-loop trajectory.}
    \label{fig:s1-p1-traj}
  \end{subfigure}%
  \begin{subfigure}[ht]{0.5\linewidth}
    \includegraphics[width=1\linewidth]{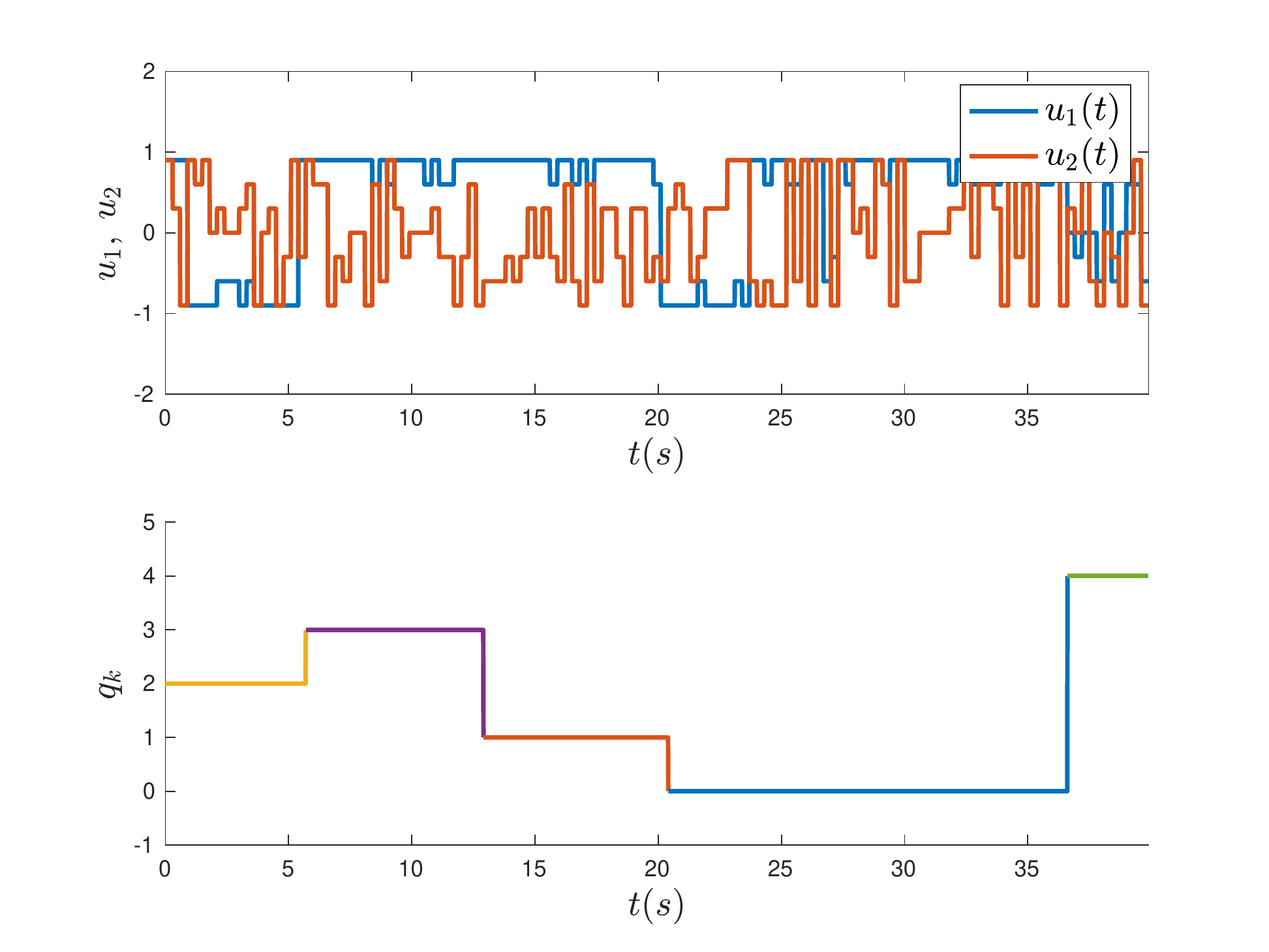}
    \caption{Time history of control inputs and DBA state.}
    \label{fig:s1-p1-u}
  \end{subfigure}
\caption{Closed-loop simulation with $x_0=(1.3,5,135^{\circ})$. The upward red and downward green triangles mark the initial and terminal states, respectively. The sections of the trajectory related to different automaton states are marked in different colors: $q_0$-blue, $q_1$-orange, $q_2$-yellow, $q_3$-purple, and $q_4$-green.}
\label{fig:traj-p1}
\end{figure}

\begin{figure}[htbp]
    \centering
    \begin{subfigure}[ht]{0.5\linewidth}
      \includegraphics[width=1\linewidth]{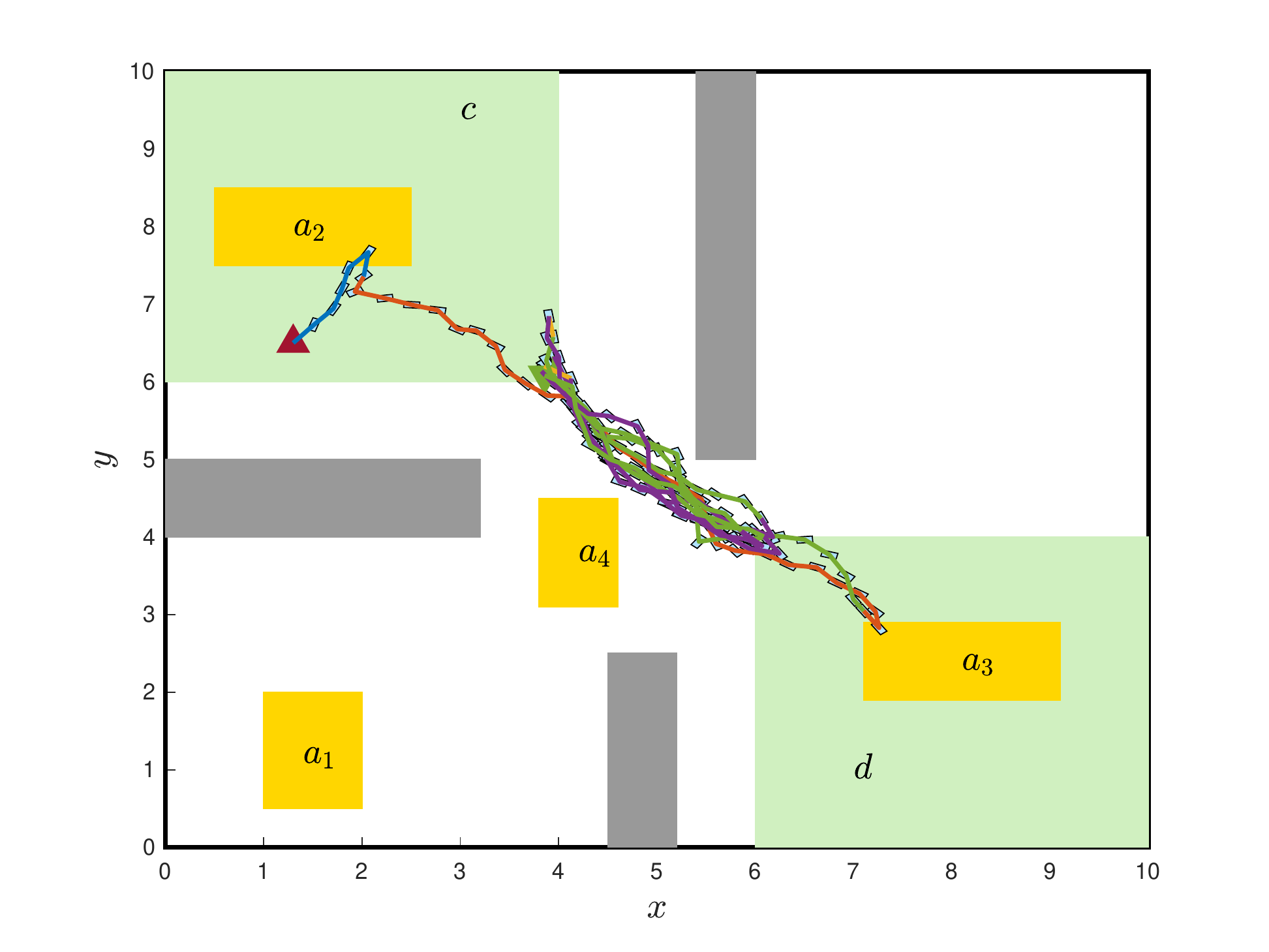}
      \caption{$x_0=(1.3,6.5, 90^\circ)$}
    \end{subfigure}%
    \begin{subfigure}[ht]{0.5\linewidth}
      \includegraphics[width=1\linewidth]{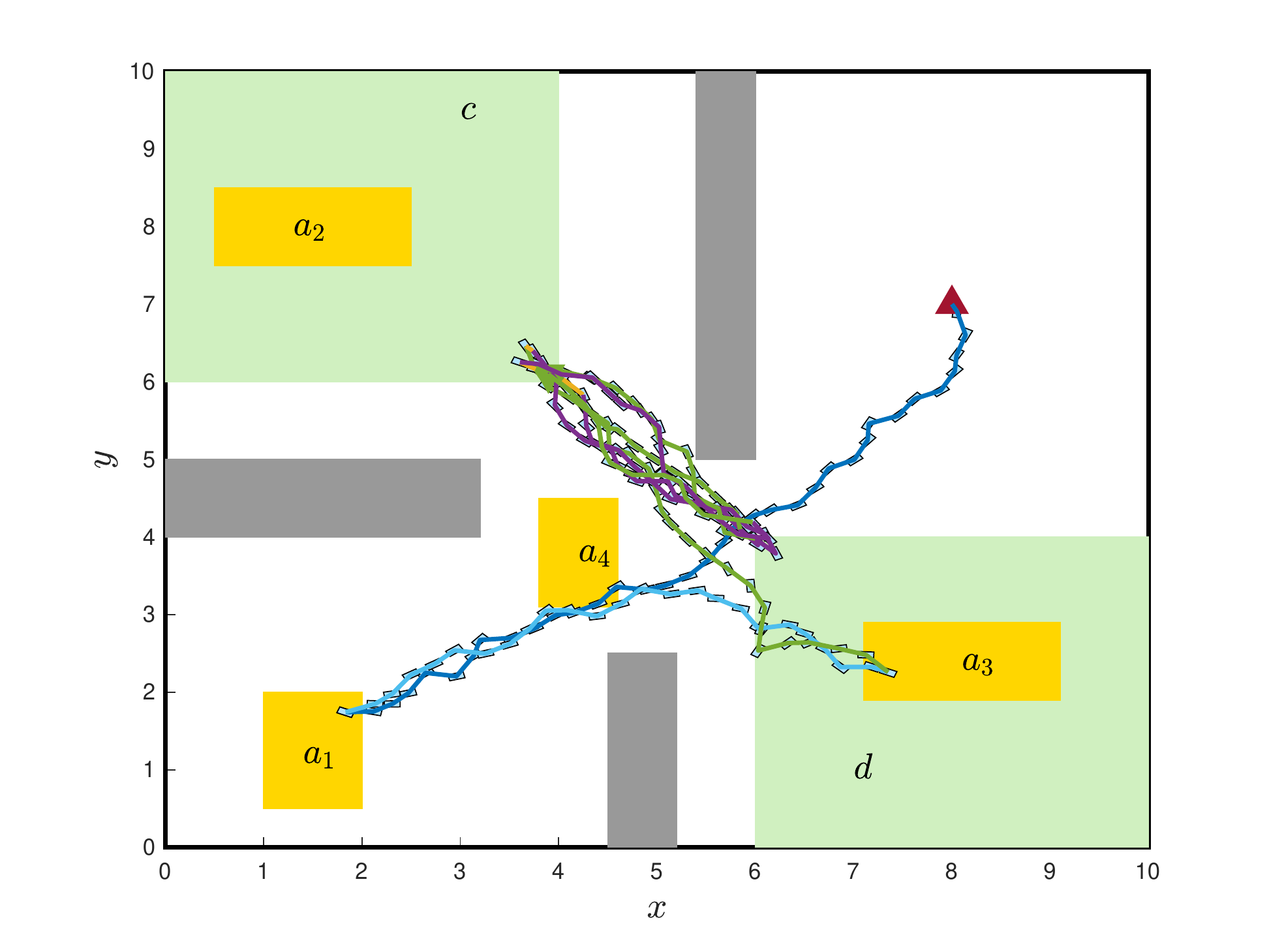}
      \caption{$x_0=(8, 7, 90^\circ)$.}
    \end{subfigure}
    \caption{The closed-loop trajectories with two different initial conditions.\vspace{-3mm}}
    \label{fig:s2-sim}
\end{figure}

Two layouts of the workspace (see Fig.~\ref{fig:s1-p1-traj} and Fig.~\ref{fig:s2-sim}) are tested, where the safety requirement $\varphi_s=\always\neg o$ is imposed. With the first layout, we aim to design controllers w.r.t. three specifications: $\varphi_1$ in Example \ref{eg:demo}, a generalized B\"uchi formula $\varphi_2=\bigwedge_{i=1}^{3} \always\eventually a_i$, and a generalized reachability formula $\varphi_3=\bigwedge_{i=1}^{3}\eventually a_i$.

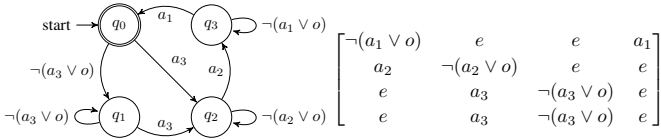
\begin{figure}[htbp]
  \centering
  \begin{minipage}{0.5\columnwidth}
    \resizebox{1\linewidth}{!}{%
    \begin{tikzpicture}[->,>=stealth',auto,node distance=2cm,semithick,baseline=-30pt]
      \node[state,initial,accepting] (v_0) {$q_0$};
      \node[state] (v_3) [below of=v_0] {$q_1$};
      \node[state] (v_2) [right of=v_0] {$q_3$};
      \node[state] (v_1) [below of=v_2] {$q_2$};
      \path (v_0) edge [bend right] node [left] {$\neg (a_3\vee o)$} (v_3)
      edge node {$a_3$} (v_1)
      (v_1) edge [loop right] node {$\neg (a_2\vee o)$} (v_1)
      edge [bend right] node {$a_2$} (v_2)
      (v_2) edge [loop right] node {$\neg (a_1\vee o)$} (v_0)
      edge [bend right] node {$a_1$} (v_0)
      (v_3) edge [bend right] node {$a_3$} (v_1)
      edge [loop left] node {$\neg (a_3\vee o)$} (v_3);
    \end{tikzpicture}}
  \end{minipage}%
  \resizebox{0.5\columnwidth}{!}{%
  \begin{minipage}{0.5\columnwidth}
  \begin{align*}
    \begin{bmatrix}
      \neg (a_1\vee o) & e & e & a_1\\
      a_2 & \neg (a_2\vee o) & e & e\\
      e & a_3 & \neg (a_3\vee o) & e\\
      e & a_3 & \neg (a_3\vee o) & e\\
    \end{bmatrix}
  \end{align*}
  \end{minipage}}
  \caption{The DBA of $\varphi_2$ and $\M_{\varphi_2}$.}
  \label{fig:gb}
\end{figure}

The formula $\varphi_1$ specifies the order of areas that the vehicle has to visit: $a_1\to a_2\to a_3\to\neg a_2\to a_1$. To satisfy $\varphi_2$, the vehicle is expected to visit three isolated areas labeled by $a_1$, $a_2$ and $a_3$ infinitely often. The DBA of $\varphi_2$ (Fig.~\ref{fig:gb}) itself is an SCC, and thus pre-processing is not needed. The transition matrix $\M_{\varphi_2}$is obtained by arranging the states in the order $q_3q_2q_1q_0$.

An equivalent DBA of $\varphi_3$ is shown in Fig.~\ref{fig:greach}), which contains no non-trivial SCC. The paths that lead to the accepting node $q_0$ enumerate the orders of visiting $a_{1,2,3}$. A topological sort gives the order of DBA nodes for $\M_{\varphi_3}$: $q_3q_2q_6q_7q_4q_5q_1q_0$.

  \begin{figure}[htbp]
    \centering
    \begin{minipage}{1\columnwidth}
    \centering
    \resizebox{0.9\linewidth}{!}{%
    \begin{tikzpicture}[->,>=stealth',auto,node distance=2.5cm,semithick]
      \node[state,initial,initial where=left] (v_3) {$q_3$};
      \node[state] (v_6) [above right of=v_3] {$q_6$};
      \node[state] (v_2) [right of=v_3] {$q_2$};
      \node[state] (v_7) [below right of=v_3] {$q_7$};
      \node[state] (v_4) [right of=v_2] {$q_4$};
      \node[state,accepting] (v_0) [right of=v_4] {$q_0$};
      \node[state] (v_5) [below left of=v_0] {$q_5$};
      \node[state] (v_1) [above left of=v_0] {$q_1$};
      \path (v_3) edge node {$a_2$} (v_6)
      edge node {$a_3$} (v_7)
      edge node {$a_1$} (v_2)
      edge [loop below] node [below] {$p_0$} (v_3)
      (v_2) edge node [above] {$a_2$} (v_1)
      edge node {$a_3$} (v_5)
      edge [loop above] node {$p_3$} (v_3)
      (v_7) edge node {$a_2$} (v_4)
      edge node {$a_1$} (v_5)
      edge [loop left] node [left] {$p_1$} (v_7)
      (v_6) edge node [above] {$a_3$} (v_4)
      edge node {$a_1$} (v_1)
      edge [loop left] node {$p_2$} (v_6)
      (v_1) edge node {$a_3$} (v_0)
      edge [loop right] node {$\neg(a_3\vee o)$} (v_1)
      (v_4) edge node {$a_1$} (v_0)
      edge [loop above] node [right] {$\neg(a_1\vee o)$} (v_4)
      (v_5) edge node {$a_2$} (v_0)
      edge [loop right] node {$\neg(a_2\vee o)$} (v_5)
      (v_0) edge [loop right] node {$\top$} (v_0);
    \end{tikzpicture}}
    \end{minipage}
    \resizebox{0.9\linewidth}{!}{%
    \begin{minipage}{0.5\columnwidth}
    \begin{align*}
    \M_{\varphi_3}=
    \begin{bmatrix}
      p_0&a_1&a_2&a_3 &e&e&e&e\\
      e&p_3&e&e&e&a_3&a_2&e\\
      e&e&p_2&e&a_3&e&a_1&e\\
      e&e&e&p_1&a_2&a_1&e&e\\
      e&e&e&e&\neg(a_1\vee o)&e&e&a_1\\
      e&e&e&e&e&\neg(a_2\vee o)&e&a_2\\
      e&e&e&e&e&e&\neg(a_3\vee o)&a_3\\
      e&e&e&e&e&e&e&\top\\
    \end{bmatrix}
    \end{align*}
    \end{minipage}
    }
    \caption{The DBA $\varphi_3$ with $p_0=\neg (a_1\vee a_2\vee a_3\vee o)$, $p_1=\neg (a_1\vee a_2\vee o)$, $p_2=\neg (a_1\vee a_3\vee o)$, $p_3=\neg (a_2\vee a_3\vee o)$.}
    \label{fig:greach}
  \end{figure}

\begin{figure}[htbp]
  \centering
  \begin{subfigure}[htbp]{0.5\linewidth}
    \includegraphics[width=1\linewidth]{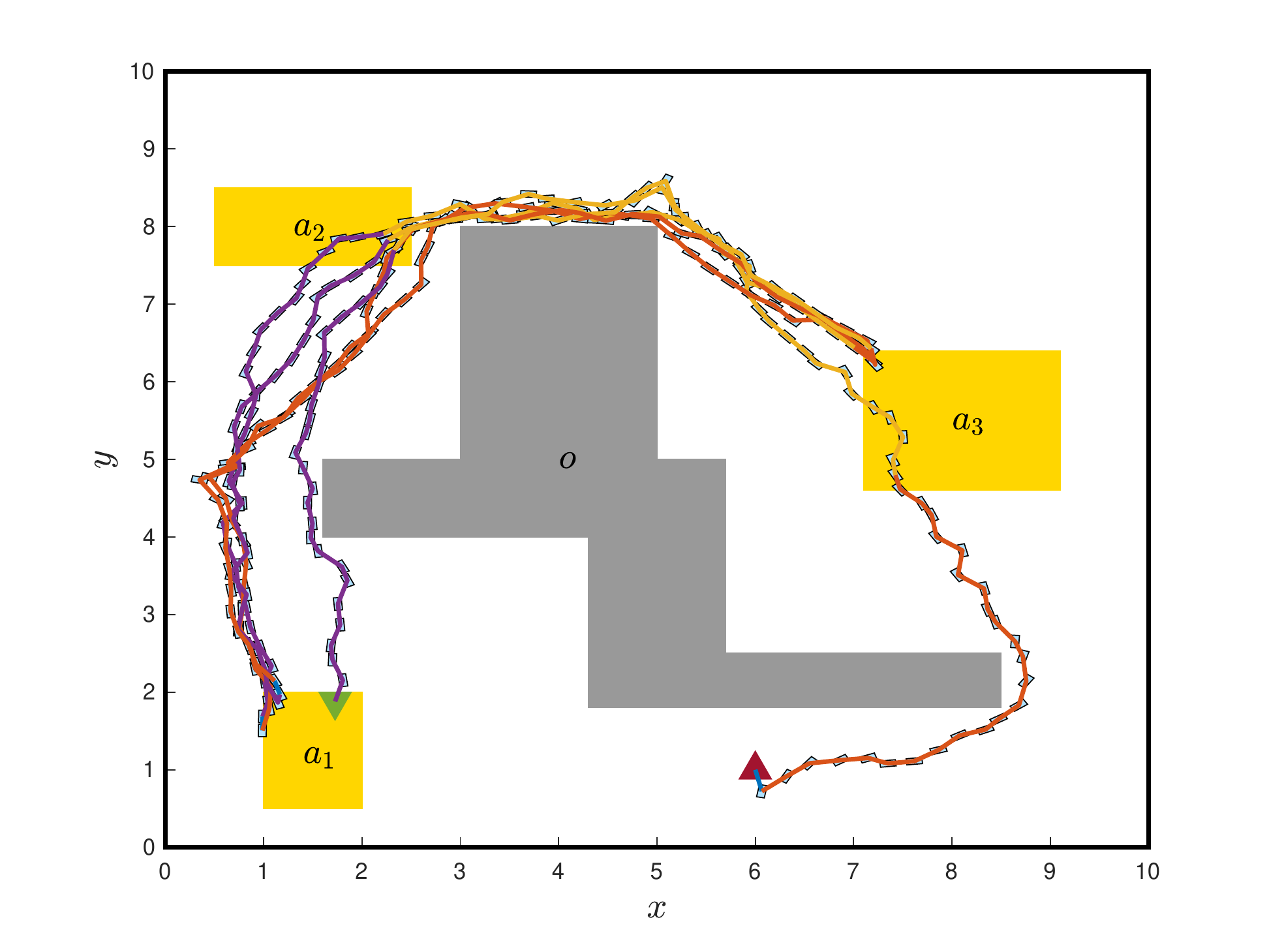}
    \caption{$x_0=(6,1,90^{\circ})$ for $\varphi_2$}
  \end{subfigure}%
  \begin{subfigure}[ht]{0.5\linewidth}
    \includegraphics[width=1\linewidth]{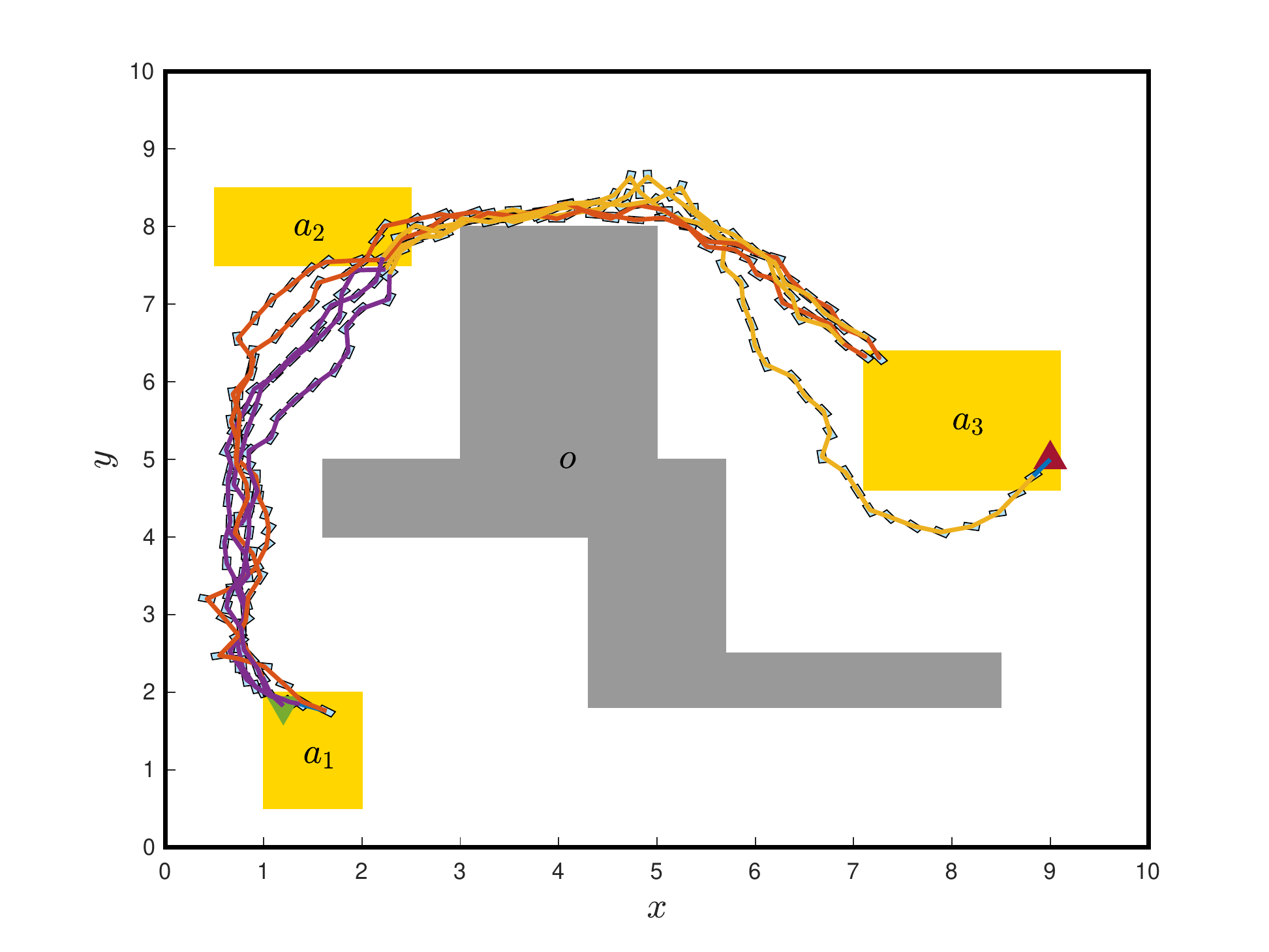}
    \caption{$x_0=(9.0,5.0,45^{\circ})$ for $\varphi_2$}
  \end{subfigure}
  \begin{subfigure}[ht]{0.5\linewidth}
    \includegraphics[width=1\linewidth]{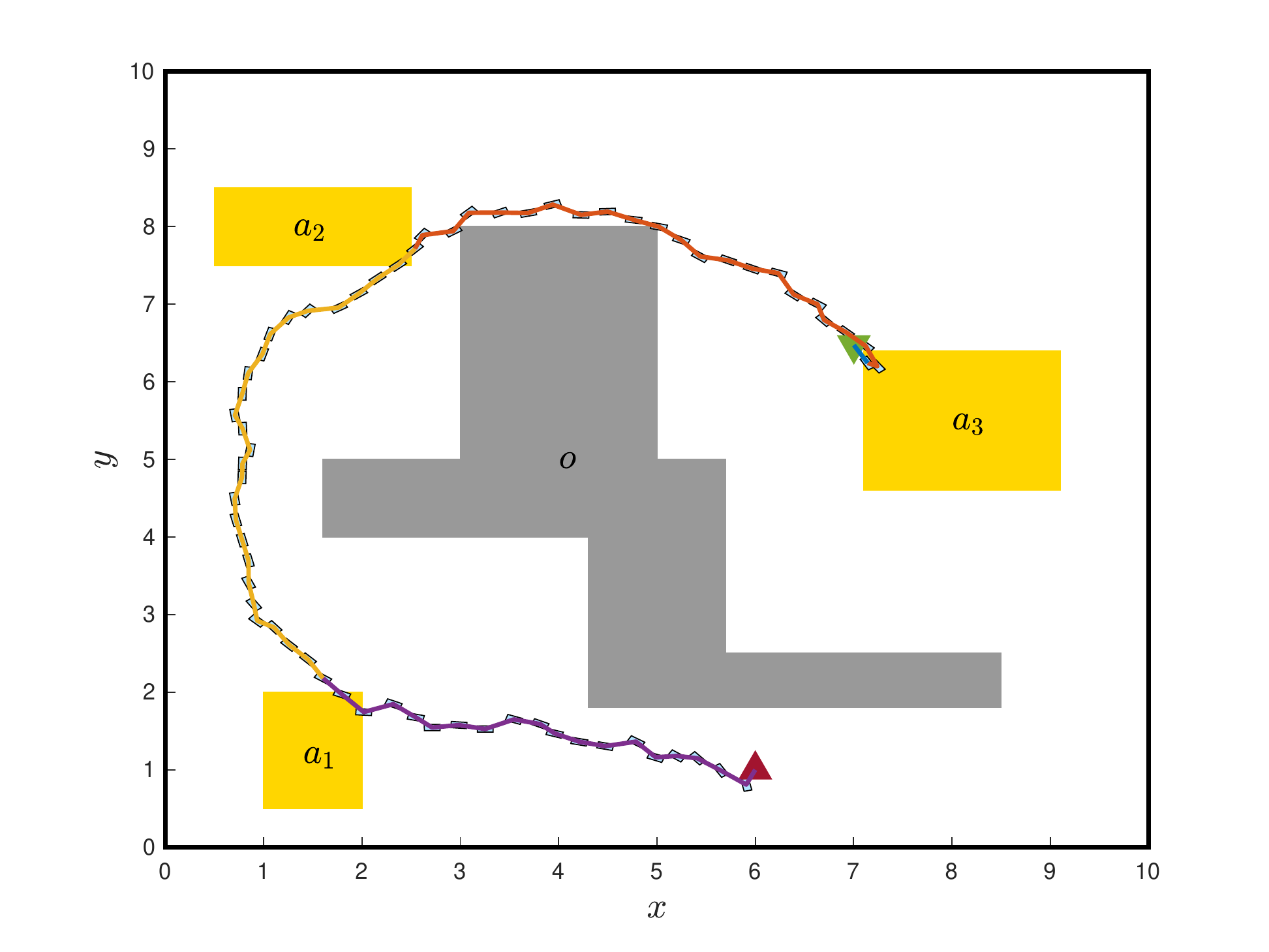}
    \caption{$x_0=(6,1,90^{\circ})$ for $\varphi_3$}
  \end{subfigure}%
  \begin{subfigure}[ht]{0.5\linewidth}
    \includegraphics[width=1\linewidth]{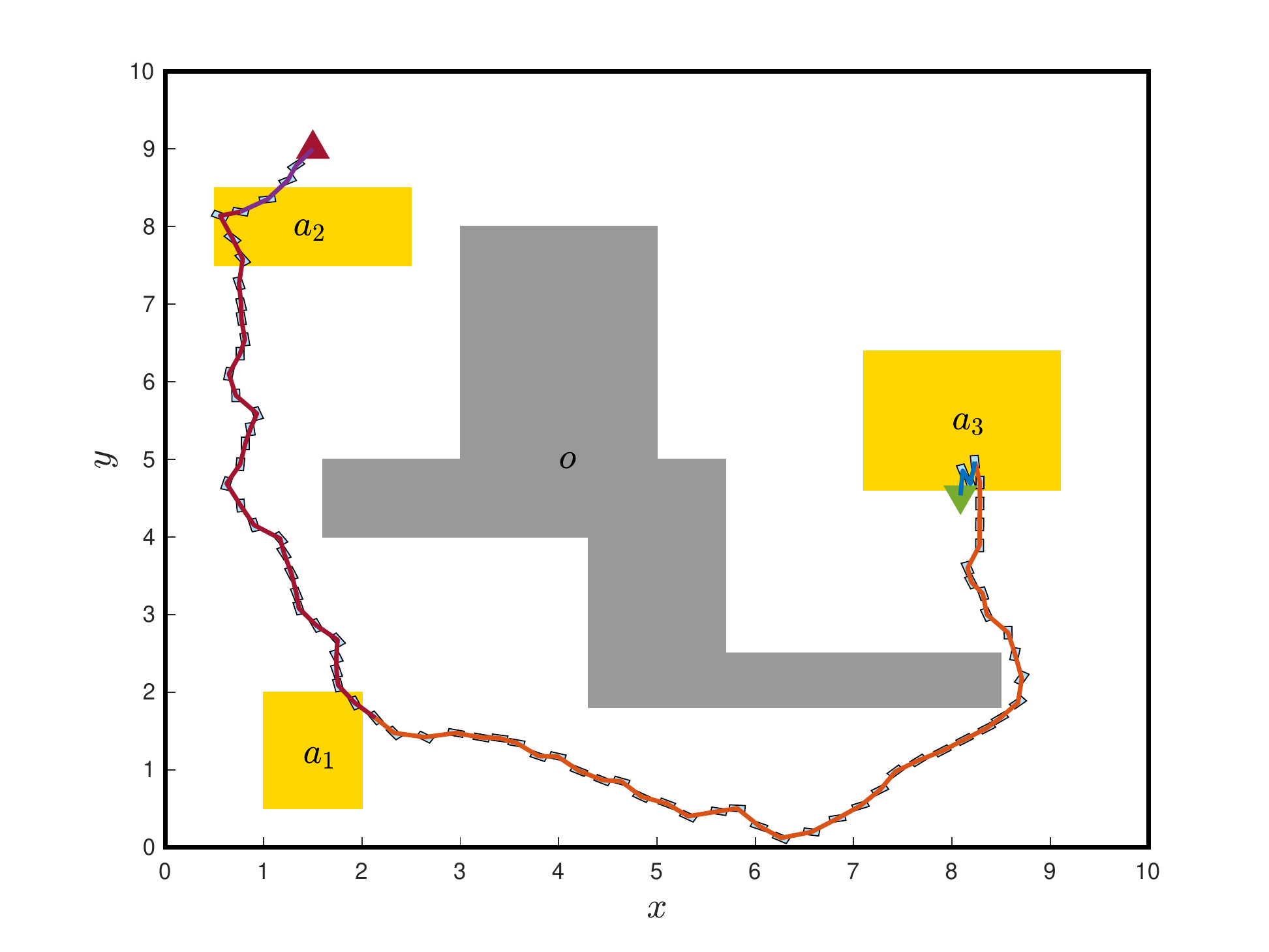}
    \caption{$x_0=(1.5,9.0,45^{\circ})$ for $\varphi_3$}
  \end{subfigure}
  \caption{Closed-loop trajectories from 4 different initial conditions that meets $\varphi_2$ ((a), (b)) and $\varphi_3$ ((c), (d)).\vspace{-3mm}}
  \label{fig:4initials}
\end{figure}

Fig.~\ref{fig:traj-p1} shows the controlled trajectory of the mobile robot for $\varphi_1$ from an initial state $x_0=(1.3,5,135^{\circ})$. The DBA in Fig.~\ref{fig:robotltl} starts from node $q_2$, and thus the robot turns back to the area $L^{-1}(a_1)$ at the beginning to trigger the transition in the DBA that leads to node $q_3$. From the trajectory and the change of DBA states, we can see that the specification $\varphi_1$ is fulfilled without collision.
Simulations of closed-loop trajectories with different initial conditions for $\varphi_{2,3}$ are shown in Fig.~\ref{fig:4initials}. 

\begin{figure}[htbp]
    \centering
    \begin{minipage}{0.8\columnwidth}
    \centering
    \resizebox{1.\linewidth}{!}{%
    \begin{tikzpicture}[->,>=stealth',auto,node distance=2.2cm,semithick]
        \node[state,initial,initial where=left] (v_0) {$q_0$};
        \node[state] (v_5) [right of=v_0] {$q_5$};
        \node[state] (v_1) [above of=v_5] {$q_1$};
        \node[state] (v_4) [right of=v_5] {$q_4$};
        \node[state,accepting] (v_2) [above right of=v_4] {$q_2$};
        \node[state] (v_3) [right of=v_4] {$q_3$};
        \path (v_0) edge [bend left] node {$a_2$} (v_1)
        edge node {$l_3$} (v_5)
        edge [loop below] node [below] {$l_1$} (v_0)
        (v_1) edge [loop above] node [above] {$l_2$} (v_1)
        edge node {$a_3\wedge d$} (v_4)
        edge node {$a_1$} (v_5)
        edge node {$a_4$} (v_0)
        (v_5) edge node {$a_3\wedge d$} (v_4)
        edge [loop below] node {$\neg (a_3\vee o)$} (v_5)
        (v_4) edge [bend left] node {$c$} (v_2)
        edge [loop below] node {$\neg (c\vee o)$} (v_4)
        (v_2) edge node {$d$} (v_4)
        edge node {$\neg (d\vee o)$} (v_3)
        (v_3) edge [loop below] node {$\neg (d\vee o)$} (v_3)
        edge node {$d$} (v_4);
    \end{tikzpicture}}
    \end{minipage}
    \resizebox{0.8\columnwidth}{!}{%
    \begin{minipage}{0.5\columnwidth}
    \begin{align*}
        \M_{\varphi_4}=\left[
        \begin{array}{ccc:ccc}
          l_1 & a_2 & l_3 & e & e & e\\
          a_4 & l_2 & a_1 & e & a_3\wedge d & e\\
          e & e & \neg (a_3\vee o) & e & a_3\wedge d & e\\\hdashline
          e & e & e &\neg (d\vee o) & d & e\\
          e & e & e & e & \neg (c\vee o) & c\\
          e & e & e & \neg (d\vee) & d & e
        \end{array}\right]
    \end{align*}
    \end{minipage}
    }
    \caption{The DBA of $\varphi_4$ with $l_1=\neg o\wedge (a_4\vee(\neg a_1\wedge\neg a_2))$, $l_2=\neg (o\vee a_1\vee a_4\vee a_3)$ and $a_1\wedge\neg (a_3\vee o)$. The topological sort gives the order $q_0q_1q_5q_3q_4q_2$ for $\M_{\varphi_4}$.}
    \label{fig:phi4}
\end{figure}
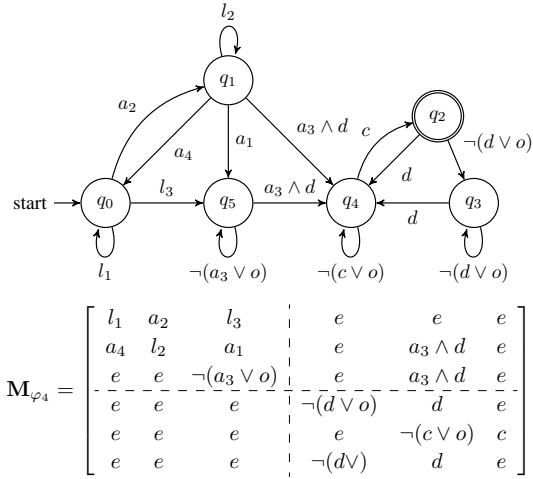

With the second layout, the task for the robot is $\varphi_4=\eventually(a_1\wedge\eventually a_3)\vee\eventually(a_2 \wedge(\neg a_4 \until a_3))\wedge \always\eventually c\wedge\always\eventually d$.
To follow $\varphi_4$, the robot not only needs to visit areas $c$ and $d$ repeatedly, but also has to go to $a_1$ then $a_3$ or $a_2$ then $a_3$ while avoiding $a_4$. The corresponding DBA shown in Fig.~\ref{fig:phi4} contains two SCCs, which indicates a matrix decomposition in the form of (\ref{eq:Mblocks}) and the pre-processing can be used to reduce the time for control synthesis.

The closed-loop simulation results with initial conditions $x_0=(1.3,6.5,90^\circ)$ and $x_0=(8, 7, 90^\circ)$ are shown in Fig.~\ref{fig:s2-sim}. 

\subsection{Experimental time and memory efficiency}
We compare the proposed method with abstraction-based methods for time and space performance evaluation. Implementations of both methods are submitted to the high performance computing (HPC) cluster B\'eluga as sequential tasks. All the jobs are run on a node with Intel Xeon(R) Gold 6148 processor @\SI{2.4}{\GHz}. Each node is allocated Maximum \SI{750}{\giga\byte} of memory.
Table~\ref{tab:examplesummary} lists system dimensions and sizes of specifications of the three cases in the previous sections.
\begin{table}[htbp]
    \centering
    \caption{A summary of examples. $n$: the state dimension; $m$: the input dimension; $\abs{Q}$: the number of nodes in the DBA; $\abs{r}$: the number of edges in the DBA.
    }
    \begin{tabular}{llllll}
        \toprule
        Cases & $n$ & $m$ & Specs & $\abs{Q}$ & $\abs{r}$\\
        \hline
        \multirow{1}{*}{Engine} & 2 & 2 & \multirow{1}{*}{$\varphi_{\rm rs}$} & 3 & 5\\
        \multirow{1}{*}{SCARA} & 4 & 2 & \multirow{1}{*}{$\varphi_{\rm gb}$} & 3 & 8\\
        \multirow{4}{*}{Vehicle} & \multirow{4}{*}{3} & \multirow{4}{*}{2} & \multirow{1}{*}{$\varphi_1$} & \multirow{1}{*}{5} & \multirow{1}{*}{10}\\
        & & & \multirow{1}{*}{$\varphi_2$} & \multirow{1}{*}{4} & \multirow{1}{*}{8}\\
        & & & \multirow{1}{*}{$\varphi_3$} & \multirow{1}{*}{8} & \multirow{1}{*}{20}\\
        & & & \multirow{1}{*}{$\varphi_4$} & \multirow{1}{*}{6} & \multirow{1}{*}{14}\\
        \bottomrule
    \end{tabular}
    \label{tab:examplesummary}
\end{table}

The critical data, such as time and memory consumption, of control synthesis by using the proposed specification-guided and abstraction-based methods are presented in Table~\ref{tab:performanceresult}. In all cases, the proposed method outperforms abstraction-based methods in terms of memory efficiency while its performance in computational time varies among examples. For the cases with high partition precision requirement, the proposed method shows more advantage since memory consumption becomes a bottleneck for abstraction-based method and the time consumption for both methods are close.

For the SCARA manipulator example, the results by using abstraction-based methods are not available since more than 750GB memory is required for control synthesis using abstraction-based methods. In the Moore-Greitzer engine example, to perform control synthesis by using abstraction-based methods w.r.t. the reach-avoid-stay specification $\varphi_{\rm rs}$, we use a sound algorithm (but not complete) since $\varphi_{\rm rs}$ cannot be translated into a DBA. As we have discussed in Section~\ref{sec:full+soundness}, Algorithm~\ref{alg:buchigame} can be used directly. To obtain a non-empty winning set, a precision parameter $\varepsilon\geq \num{1.8e-4}$ is required. The proposed method is more than 300 times more efficient than abstraction-based methods in memory consumption, even though the run time is around 4 times of that of abstraction-based methods.
\begin{table*}[htbp]
    \centering
    \caption{Performance comparison with abstraction-based methods. ``Time1" and ``Time2" are for the run times by using the proposed method with and without pre-processing, respectively. ``Win\%"= winning set percentage of the state space. $N_\X$ for the specification-guided method is the average number of partitions among $\abs{Q}$ returned $\sys$-domains.}
    \resizebox{\linewidth}{!}{%
    \begin{tabular}{@{}lllllllllllll@{}}
        \toprule
        \multirow{2}{*}{Cases} & \multirow{2}{*}{Specs} & \multirow{2}{*}{$\varepsilon$} & \multicolumn{5}{c}{Specification-guided} & \multicolumn{5}{c}{Abstraction-based}\\
        \cmidrule(rl){4-8} \cmidrule(rl){9-13}
        & & & $N_\X$ & Win\% & Time1(s) & Time2(s) & Mem(GB) & $N_\X$ & $N_R$ & Win\% & Time(s) & Mem(GB) \\
        \hline
        \multirow{2}{*}{Engine} & \multirow{1}{*}{$\varphi_{\rm rs1}$} & \multirow{2}{*}{\num{1.8e-4}} &279200 &99.64 &21883 &N/A &0.357 &\multirow{2}{*}{1236544} &1670780774 &99.63 &5142.08 &117.58 \\
        & \multirow{1}{*}{$\varphi_{\rm rs2}$} & &220493 &99.64 &16814.4 &N/A &0.289 & &1670711172 &99.63 &3909.23 &117.54 \\
        \multirow{1}{*}{SCARA} & \multirow{1}{*}{$\varphi_{\rm gb}$} & 0.05 &2640629 &49.31 &41405.7& N/A &5.45 & N/A & N/A & N/A & N/A & $>750$\\
        \multirow{4}{*}{Vehicle} & \multirow{1}{*}{$\varphi_1$} & \multirow{4}{*}{0.2} &222662 &84.32 &347.207 &308.527 & 0.435 &\multirow{4}{*}{93636} &\multirow{4}{*}{26888839} &80.09 &12.569 &3.01\\
        & \multirow{1}{*}{$\varphi_2$} & &209002 &84.32 &556.5 &N/A & 0.33 & & &80.09 &11.712 &2.97\\
        & \multirow{1}{*}{$\varphi_3$} & &240463 &84.83 &576.296 &N/A &0.727 & & &80.09 &21.837 &4.08 \\
        & \multirow{1}{*}{$\varphi_4$} & &241380 &91.45 &852.76 &521.45 &0.556 & & &87.21 &18.373 &3.28 \\
        \bottomrule
    \end{tabular}
    }
    \label{tab:performanceresult}
\end{table*}

In the mobile robot motion planning example, by using a precision parameter as coarse as $\varepsilon=0.2$ for each specification, more than \SI{80}{\percent} area of the state space can be controlled to satisfy the specification by using both methods (specific data are shown in columns ``Win\%" in Table~\ref{tab:performanceresult}). The winning sets obtained by the proposed method are larger than the ones by abstraction-based methods. This is because the system state space is partitioned based on the specified regions in the proposed method while the partition is done uniformly irrespective of the specification in abstraction-based methods. Additionally, the precision $\varepsilon$ is used as the stopping criterion of the bisection in Algorithm~\ref{alg:cpre}. As a result, the smallest size of the interval is generally smaller than $\varepsilon$ and, in the worst case, could be close to $\varepsilon/2$.

\begin{figure}[htbp]
    \centering
    \includegraphics[width=\columnwidth]{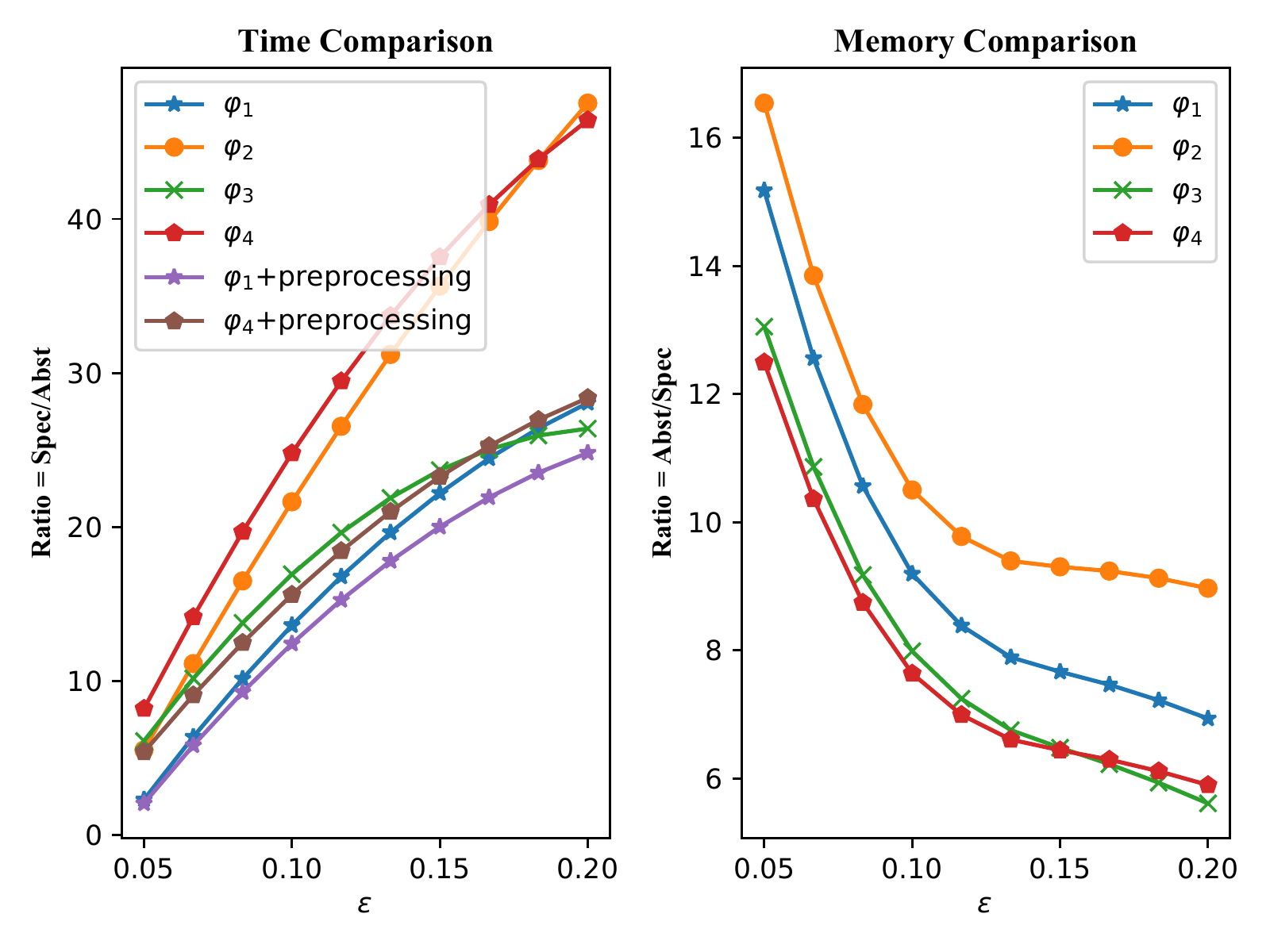}
    \caption{Time and memory comparison of the proposed specification-guided and abstraction-based methods.}
    \label{fig:comparison}
\end{figure}

By using precision $\varepsilon=0.2$, which induces a discrete abstraction of small size, abstraction-based methods shows obvious advantage in time efficiency than the proposed method. To experimentally analyze how these two methods performs w.r.t. the precision parameter $\varepsilon$, we also study the mobile robot motion planning under different precisions, and Fig.~\ref{fig:comparison} shows the comparison results in terms of time and memory efficiency. As the precision goes higher, the proposed method can save as high as 16 times of the memory usage of abstraction-based methods while its computational time is only around 4 times of abstraction-based methods. Considering both time and memory performance, the proposed method has more advantage over abstraction-based methods when a small precision parameter is necessary in order to successfully find a control strategy.
We can also observe from Table~\ref{tab:performanceresult} and Fig.~\ref{fig:comparison} that run times for specifications $\varphi_1$ and $\varphi_4$, where pre-processing is applicable, are improved by around \SI{10}{\percent} and \SI{37}{\percent}, respectively.

\section{Conclusion}\label{sec:conclude}

This work deals with control synthesis problems for nonlinear systems w.r.t. the class of LTL formulas that can be translated to DBA. The contributions are threefold. Firstly, we gave a precise characterization of the winning set (w.r.t. a DBA specification) on the continuous state space of a nonlinear system, instead of on an approximated discrete state space. Secondly, based on such a characterization, we adopted an adaptive set approximation scheme by using interval computation for practical and efficient control synthesis. We showed that, without additional assumptions, the proposed control synthesis algorithm is sound for any LTL formula and complete for synthesizing w.r.t. the DBA specifications that can be realized for the perturbed system. The proposed method is also proved to be memory efficient since the construction of abstractions is avoided. Thirdly, we proposed a pre-processing procedure before control synthesis, which can reduce the computational cost by utilizing the structure in the given LTL formula. Furthermore, we demonstrated that the proposed method can effectively solve the motion planning problem for manipulators and mobile robots with non-convex constraints on their workspaces. Our future work will focus on the extension of the current result to the full set of LTL formulas as well as developing a parallel computation algorithm to improve the time efficiency.

\appendices



\ifCLASSOPTIONcaptionsoff
  \newpage
\fi



\bibliographystyle{IEEEtran}
\bibliography{dba}

\begin{thebibliography}{10}
\providecommand{\url}[1]{#1}
\csname url@samestyle\endcsname
\providecommand{\newblock}{\relax}
\providecommand{\bibinfo}[2]{#2}
\providecommand{\BIBentrySTDinterwordspacing}{\spaceskip=0pt\relax}
\providecommand{\BIBentryALTinterwordstretchfactor}{4}
\providecommand{\BIBentryALTinterwordspacing}{\spaceskip=\fontdimen2\font plus
\BIBentryALTinterwordstretchfactor\fontdimen3\font minus
  \fontdimen4\font\relax}
\providecommand{\BIBforeignlanguage}[2]{{%
\expandafter\ifx\csname l@#1\endcsname\relax
\typeout{** WARNING: IEEEtran.bst: No hyphenation pattern has been}%
\typeout{** loaded for the language `#1'. Using the pattern for}%
\typeout{** the default language instead.}%
\else
\language=\csname l@#1\endcsname
\fi
#2}}
\providecommand{\BIBdecl}{\relax}
\BIBdecl

\bibitem{BeltaBook2017}
C.~Belta, B.~Yordanov, and E.~{Aydin Gol}, \emph{{Formal Methods for
  Discrete-Time Dynamical Systems}}.\hskip 1em plus 0.5em minus 0.4em\relax
  Springer International Publishing, 2017.

\bibitem{Bhatia2011}
``Motion planning with complex goals,'' \emph{IEEE Robot. Autom. Mag.},
  vol.~18, no.~3, pp. 55--64, 2011.

\bibitem{Plaku2015}
E.~Plaku and S.~Karaman, ``Motion planning with temporal-logic specifications:
  Progress and challenges,'' \emph{AI Commun.}, vol.~29, no.~1, pp. 151--162,
  2015.

\bibitem{Zielonka1998}
W.~Zielonka, ``Infinite games on finitely coloured graphs with applications to
  automata on infinite trees,'' \emph{Theor. Comput. Sci.}, vol. 200, no. 1-2,
  pp. 135--183, 1998.

\bibitem{Kupferman2001}
O.~Kupferman and M.~{Y. Vardi}, ``Model checking of safety properties,''
  \emph{Form. Methods Syst. Des.}, vol.~19, no.~3, pp. 291--314, 2001.

\bibitem{Tabuada2009verification}
P.~Tabuada, \emph{{Verification and Control of Hybrid Systems: A Symbolic
  Approach}}.\hskip 1em plus 0.5em minus 0.4em\relax Springer Science \&
  Business Media, 2009.

\bibitem{ZamaniPMT12}
M.~Zamani, G.~Pola, M.~M. Jr., and P.~Tabuada, ``Symbolic models for nonlinear
  control systems without stability assumptions,'' \emph{IEEE Trans. Automat.
  Contr.}, vol.~57, no.~7, pp. 1804--1809, 2012.

\bibitem{liu2016finite}
J.~Liu and N.~Ozay, ``Finite abstractions with robustness margins for temporal
  logic-based control synthesis,'' \emph{Nonlinear Anal. Hybrid Syst.},
  vol.~22, pp. 1--15, 2016.

\bibitem{Reissig2016}
G.~Reissig, A.~Weber, and M.~Rungger, ``Feedback refinement relations for the
  synthesis of symbolic controllers,'' \emph{IEEE Trans. Automat. Contr.},
  vol.~62, no.~4, pp. 1781 -- 1796, 2017.

\bibitem{liu2017robust}
J.~Liu, ``Robust abstractions for control synthesis: Robustness equals
  realizability for linear-time properties,'' in \emph{Proc. of HSCC}, 2017,
  pp. 101--110.

\bibitem{MooreBook66}
R.~E. Moore, \emph{{Interval Analysis}}.\hskip 1em plus 0.5em minus 0.4em\relax
  Prentice-Hall, 1966.

\bibitem{chatterjee2008algorithms}
K.~Chatterjee, T.~A. Henzinger, and N.~Piterman, ``Algorithms for {B}\"uchi
  games,'' in \emph{Proceedings of 3rd Workshop on Games in Design and
  Verification}, 2006.

\bibitem{chatterjee2012n}
K.~Chatterjee and M.~Henzinger, ``An $o(n^2)$ time algorithm for alternating
  {B}{\"u}chi games,'' in \emph{Proceedings of the twenty-third annual ACM-SIAM
  symposium on Discrete Algorithms}.\hskip 1em plus 0.5em minus 0.4em\relax
  SIAM, 2012, pp. 1386--1399.

\bibitem{de2001symbolic}
L.~De~Alfaro, T.~A. Henzinger, and R.~Majumdar, ``Symbolic algorithms for
  infinite-state games,'' in \emph{International Conference on Concurrency
  Theory}.\hskip 1em plus 0.5em minus 0.4em\relax Springer, 2001, pp. 536--550.

\bibitem{Wolff2013icirs}
E.~M. Wolff, U.~Topcu, and R.~M. Murray, ``Automaton-guided controller
  synthesis for nonlinear systems with temporal logic,'' in \emph{Proc. of
  IROS}, 2013, pp. 4332--4339.

\bibitem{AydinGol2014}
E.~{Aydin Gol}, M.~Lazar, and C.~Belta, ``Language-guided controller synthesis
  for linear systems,'' \emph{IEEE Trans. Automat. Contr.}, vol.~59, no.~5, pp.
  1163--1176, 2014.

\bibitem{Kloetzerhscc08}
M.~Kloetzer and C.~Belta, ``Dealing with nondeterminism in symbolic control,''
  in \emph{Proc. of HSCC}, 2008, pp. 287--300.

\bibitem{Li2016}
Y.~Li and J.~Liu, ``Invariance control synthesis for switched nonlinear
  systems: An interval analysis approach,'' \emph{IEEE Trans. Automat. Contr.},
  vol.~63, no.~7, pp. 2206--2211, 2018.

\bibitem{Li2020Reachstay}
------, ``Robustly complete synthesis of memoryless controllers for nonlinear
  systems with reach-and-stay specifications,'' \emph{IEEE Trans. Automat.
  Contr.}, vol.~66, no.~3, pp. 1199--1206, 2021.

\bibitem{Zibaeenejad2019}
M.~H. Zibaeenejad and J.~Liu, ``Auditor product and controller synthesis for
  non-deterministic transition systems with practical ltl specifications,''
  \emph{IEEE Trans. Automat. Contr.}, pp. 1--1, 2019.

\bibitem{Kim2018}
E.~S. Kim, M.~Arcak, and M.~Zamani, ``Constructing control system abstractions
  from modular components,'' in \emph{Proc. of HSCC}, 2018, pp. 137--146.

\bibitem{Hsu2018}
K.~Hsu, R.~Majumdar, K.~Mallik, and A.-K. Schmuck, ``Multi-layered
  abstraction-based controller synthesis for continuous-time systems,'' in
  \emph{Proc. of HSCC}, 2018, pp. 120--129.

\bibitem{Khaled2019tacas}
M.~Khaled, E.~S. Kim, M.~Arcak, and M.~Zamani, ``Synthesis of symbolic
  controllers: A parallelized and sparsity-aware approach,'' in \emph{Proc. of
  TACAS}, 2019.

\bibitem{Tarjan1972scc}
R.~Tarjan, ``Depth-first search and linear graph algorithms,'' \emph{SIAM
  Journal on Computing}, vol.~1, no.~2, pp. 146--160, 1972.

\bibitem{baier2008principles}
C.~Baier, J.-P. Katoen, and K.~G. Larsen, \emph{Principles of Model
  Checking}.\hskip 1em plus 0.5em minus 0.4em\relax MIT press, 2008.

\bibitem{Gradel2002}
E.~Gr{\"{a}}del, W.~Thomas, and T.~Wilke, Eds., \emph{{Automata Logics, and
  Infinite Games}}.\hskip 1em plus 0.5em minus 0.4em\relax Springer Berlin
  Heidelberg, 2002.

\bibitem{Li2018rocs}
Y.~Li and J.~Liu, ``{ROCS}: A robustly complete control synthesis tool for
  nonlinear dynamical systems,'' in \emph{Proc. of HSCC}, 2018, pp. 130--135.

\bibitem{Li2019thesis}
Y.~Li, ``{Robustly Complete Temporal Logic Control Synthesis for Nonlinear
  Systems},'' Ph.D. dissertation, University of Waterloo, 2019.

\bibitem{piterman2006faster}
N.~Piterman and A.~Pnueli, ``Faster solutions of rabin and streett games,'' in
  \emph{21st Annual IEEE Symposium on Logic in Computer Science
  (LICS'06)}.\hskip 1em plus 0.5em minus 0.4em\relax IEEE, 2006, pp. 275--284.

\bibitem{Duret-Lutz2016}
A.~Duret-Lutz and et~al, ``Spot 2.0 --a framework for {LTL} and
  $\omega$-automata manipulation,'' in \emph{Proc. of ATVA}, 2016, pp.
  122--129.

\bibitem{Li2022}
Y.~Li and J.~Liu, ``Robustly complete synthesis of sampled-data control for
  continuous-time nonlinear systems with reach-and-stay objectives,''
  \emph{Nonlinear Anal. Hybrid Syst.}, vol.~44, p. 101170, may 2022.

\bibitem{Siciliano2010RoboticsBook}
B.~Siciliano, L.~Sciavicco, L.~Villani, and G.~Oriolo, \emph{Robotics:
  Modelling, Planning and Control}.\hskip 1em plus 0.5em minus 0.4em\relax
  Springer Publishing Company, Incorporated, 2010.

\bibitem{Murray1994ManipulatorBook}
R.~M. Murray, S.~S. Sastry, and L.~Zexiang, \emph{A Mathematical Introduction
  to Robotic Manipulation}, 1st~ed.\hskip 1em plus 0.5em minus 0.4em\relax USA:
  CRC Press, Inc., 1994.

\bibitem{AstromM08}
K.~J. Astrom and R.~M. Murray, \emph{{Feedback Systems: An Introduction for
  Scientists and Engineers}}.\hskip 1em plus 0.5em minus 0.4em\relax Princeton,
  2008.

\end{thebibliography}
%



%






\end{document}